\documentclass[reprint,superscriptaddress,longbibliography,eqsecnum,nofootinbib,amsmath,amssymb,aps,floatfix,]{revtex4-1}

\usepackage[utf8]{inputenc}
\usepackage[T1]{fontenc}
\usepackage[dvipsnames]{xcolor}
\usepackage{graphicx}
\usepackage{bm}
\usepackage[colorlinks=true,allcolors=blue]{hyperref}
\usepackage{mathrsfs}
\usepackage[utf8]{inputenc}
\usepackage{leftidx}

\newcommand{\UVA}{Department of Physics, University of Virginia, P.O.~Box 400714, Charlottesville, VA 22904-7414, USA}

\begin{document}

\title{Definitions of (super) angular momentum in asymptotically flat spacetimes: Properties and applications to compact-binary mergers}

\author{Arwa Elhashash}
\email{aze5tn@virginia.edu}
\affiliation{\UVA}%

\author{David A.~Nichols}%
\email{david.nichols@virginia.edu}
\affiliation{\UVA}

\date{\today}

\begin{abstract}
The symmetries of asymptotically flat spacetimes in general relativity are given by the Bondi-Metzner-Sachs (BMS) group, though there are proposed generalizations of its symmetry algebra.
Associated with each symmetry is a charge and a flux, and the values of these charges and their changes can characterize a spacetime. 
The charges of the BMS group are relativistic angular momentum and supermomentum (which includes 4-momentum); the extensions of the BMS algebra also include generalizations of angular momentum called ``super angular momentum.''
Several different formalisms have been used to define angular momentum, and they produce nonequivalent expressions for the charge.
It was shown recently that these definitions can be summarized in a two-parameter family of angular momenta, which we investigate in this paper.
We find that requiring that the angular momentum vanishes in flat spacetime restricts the two parameters to be equal.
If we do not require that the angular momentum agrees with a common Hamiltonian definition, then we are left with a one-parameter family of angular momenta that includes the definitions from the several different formalisms.
We then also propose a similar two-parameter family of super angular momentum.
We examine the effect of the free parameters on the values of the angular momentum and super angular momentum from nonprecessing binary-black-hole mergers. 
The definitions of angular momentum differ at a high post-Newtonian order for these systems, but only when the system is radiating gravitational waves (not before and after).
The different super-angular-momentum definitions occur at lower orders, and there is a difference in the change of super angular momentum even after the gravitational waves pass, which arises because of the gravitational-wave memory effect. 
We estimate the size of these effects using numerical-relativity surrogate waveforms and find they are small but resolvable.
\end{abstract}

\maketitle

\tableofcontents

\section{\label{sec:intro} Introduction}

The LIGO, Virgo, and KAGRA collaborations have now announced the detection of almost fifty binary-black-hole (BBH) mergers during the first three observing runs of the advanced-detector era beginning in 2015~\cite{LIGOScientific:2018mvr,Abbott:2020niy}.
There are a few ways in which these BBH mergers are characterized: for example, by the masses and spins of the individual black holes (BHs) plus the orbital elements of the binary at a given reference frequency or by the final mass and spin of the BH formed after the merger and ringdown (e.g.,~\cite{LIGOScientific:2018mvr,Abbott:2020niy}).
An alternate way to characterize asymptotically flat systems is in terms of the ``conserved'' quantities conjugate to the symmetries of asymptotically flat spacetimes and the net fluxes of these conserved quantities.
The symmetries of asymptotically flat spacetimes form the Bondi-Metzner-Sachs (BMS) group, which consists of transformations isomorphic to the Lorentz group and supertranslations (of which the four spacetime translations are a subgroup)~\cite{Bondi:1962px,Sachs:1962zza,Sachs:1962wk}.
The radiated energy and linear momentum (often expressed as a recoil velocity) being the quantities conjugate to the translation symmetries are often quoted when describing BBH mergers (see, e.g.,~\cite{Varma:2018aht} and references therein).        

The flux of angular momentum (the quantity related to Lorentz symmetries) is somewhat more subtle.
Angular momentum must be computed about an origin in flat spacetime; in terms of the symmetries that form the Poincar\'e group, this implies that a translation must be specified to identify the particular Lorentz transformation under consideration.
There is thus a four-parameter family of Lorentz transformations spanned by a basis of the spacetime translations in the Poincar\'e group.
In asymptotically flat spacetimes, this four-parameter family is enlarged to a countably infinite family of Lorentz transformations, each of which is associated with some basis element of the infinite-dimensional supertranslation subgroup in the BMS group.
In stationary spacetimes, there is a natural way to choose a ``preferred'' set of supertranslations that reduces the dependence of the angular momentum to a choice of origin as in flat spacetime (see~\cite{VanDerBurg1966,Newman:1966ub} or more recently~\cite{Flanagan:2015pxa}); however, in nonstationary solutions, there is no such natural choice, though there are several different proposals to ``fix'' the supertranslation freedom (see, e.g.,~\cite{Adamo:2009vu} for a review).
The absence of this preferred Poincar\'e group is referred to as the ``supertranslation ambigutity'' of angular momentum in asymptotically flat spacetimes, which is, in essence, a statement that angular momentum in asymptotically flat spacetimes is different from its counterpart in flat spacetimes.

This additional complexity in describing the value of angular momentum for an asymptotically flat spacetime may have contributed to it and its flux being less frequently quoted in the output of numerical-relativity (NR) simulations of merging black holes.
The six degrees of freedom in the relativistic angular momentum are often split into the three spin parts (corresponding to rotations) and three center-of-mass (CM) parts (corresponding to Lorentz boosts).
Of these six components, the most commonly given from NR simulations of BBHs are the magnitude of the final BH's spin (though this spin is most often computed from quasilocal constructions on the BH's apparent horizon rather than in terms of quantities measured at or near future null infinity~\cite{Owen:2007dya,Cook:2007wr,Lovelace:2008tw}); additional components of the angular momentum were computed in~\cite{Handmer:2016mls}, for example.

In addition to the supertranslation ambiguities, a number of different definitions of the angular momentum of an asymptotically flat spacetime were (and continue to be) used.
A nonexhaustive list of some of these definitions include one based on the Landau-Lifshitz pseudotensor for the intrinsic part of the angular momentum (in the CM frame of the source)~\cite{RevModPhys.52.299}, a definition based on constructions called ``linkages''~\cite{Geroch:1981ut}, ones inspired from twistor theory~\cite{Penrose:1982wp,Dray:1984rfa}, and those related to Hamiltonians conjugate to conserved quantities~\cite{Ashtekar:1981bq,Wald:1999wa}. 
When considered in their respective domains of validity, the different definitions of the angular momentum described above agree~\cite{Wald:1999wa,Nichols:2017rqr}.
More recently, however, new definitions of angular momenta arose from revisiting the Landau-Lifshitz formalism when not restricted to the CM frame~\cite{Bonga:2018gzr} and from considerations about soft theorems~\cite{Pasterski:2015tva} (particularly a subleading correction to Weinberg's soft theorem~\cite{Weinberg:1965nx}; see~\cite{Strominger:2017zoo} for a review).

It was pointed out in~\cite{Compere:2019gft} that these new definitions of angular momentum differ from the Hamiltonian definition of Wald and Zoupas~\cite{Wald:1999wa}.\footnote{Note that what we call the six-parameter (Lorentz-covariant) angular momentum, Comp\`ere \textit{et al}.\ in~\cite{Compere:2019gft} call the ``Lorentz charge.'' We also have different usages for how we describe the parts that correspond to the rotations and the Lorentz boosts.
We both call the part corresponding to Lorentz boosts ``center-of-mass angular momentum,'' but Comp\`ere \textit{et al}.\ call the parts corresponding to rotations simply ``angular momentum,'' whereas we refer to it as ``intrinsic'' or ``spin'' angular momentum, because it reduces to those quantities in the rest-frame of the source.}
Moreover, it was shown that the discrepancies in these definitions can be written in terms of two functions that are quadratic in the shear related to the outgoing GWs in asymptotically flat spacetimes.
The different definitions were parametrized in terms of two real coefficients multiplying these two quadratic functions, respectively, and when the coefficients equal one, the Hamiltonian definition of~\cite{Wald:1999wa} is recovered.
All members of this two-parameter family of angular momenta satisfy flux balance laws, are covariant with respect to quantities defined on 2-sphere cross sections of null infinity, and lead to the same correspondence with the subleading soft theorem~\cite{Compere:2019gft}.
This led Comp\`ere \textit{et al}.\ in~\cite{Compere:2019gft} to conclude that there was not a compelling physical reason to prefer one definition over another and to suggest that there could be a two-parameter family of self-consistent definitions of angular momentum of asymptotically flat spacetimes.
Comp\`ere \textit{et al}.\ later described in~\cite{Compere:2020lrt} the sense in which these different definitions can all be considered to be Hamiltonian definitions [which is why we take care to describe which (or whose) Hamiltonian definition of the charge is being used].

In this paper, we investigate this new two-parameter family of angular momenta in greater detail.
Ashtekar and Winicour~\cite{AshtekarWinicour1982} had a larger set of criteria that a charge at null infinity should satisfy than the conditions discussed in~\cite{Compere:2019gft}.\footnote{We thank Laurent Friedel for pointing out this reference to us.}
Among these conditions was requiring that the charges and fluxes vanish in flat spacetime.
We find that if we require the angular momentum to vanish in flat spacetime, then two of the parameters must be equal, thereby reducing the two parameters to one.
This calculation further implies that the one-parameter family of angular momenta will agree in any region of spacetime in which there is only electric-parity shear (which includes stationary solutions and some radiative solutions).
If we do not require that the angular momentum agree with the Wald-Zoupas definition, then we are left with a one-parameter definition that encompasses several other definitions used in the literature.

Ashtekar and Winicour further require that a charge agree with the Komar formula whenever there is an exact (as opposed to asymptotic) symmetry. 
The same calculation showing that the charge vanishes in flat spacetime also implies that the angular momentum will agree with the Komar formula~\cite{Komar:1958wp} in regions of vanishing electric-partiy shear (which include stationary regions); however, in regions with shear of generic parity, it is only the Wald-Zoupas charge that agrees with the Komar formula (by construction).\footnote{We thank Kartik Prabhu for making us aware of this property of the angular momentum.}
While this is arguably a compelling reason to consider only the Wald-Zoupas charge, we do not aim to settle the issue of whether there is a preferred definition of angular momentum among this one-parameter family here; rather, we explore whether the different commonly used definitions of angular momentum have significant differences for strongly gravitating and dynamical systems, such as the binary black holes, which have been measured observationally by LIGO and Virgo.
In this sense, our investigation is similar in spirit to that of~\cite{Ashtekar:2019rpv}, in which the effect of the supertranslation ambiguities on the angular momentum radiated from compact-binary coalescences was studied as a way to assess how large the effect could be for this class of sources.

With this approach in mind, for this residual one-parameter family of angular momenta, we expand the difference of the angular momentum from the Wald-Zoupas definition in terms of spin-weighted spherical-harmonic moments of the GW strain.
These difference terms involve only products of electric- and magnetic-type spherical-harmonic coefficients (unlike the flux of the Wald-Zoupas angular momentum), which is consistent with the results of~\cite{Compere:2019gft}.
This implies that the difference will vanish in stationary regions of spacetimes and nonradiative regions of spacetime with vanishing magnetic shear, though more generally, it will not vanish.
We compute the time-dependent difference terms for nonspinning BBH mergers, and we find that they are small compared to the total radiated angular momentum.

In addition to the BMS group, there are two different proposals for larger symmetry groups or algebras of asymptotically flat spacetimes.
The first, due to Barnich and Troessaert~\cite{Barnich:2009se,Barnich:2010eb,Barnich:2011mi}, considers all the conformal Killing vectors of the 2-sphere, rather than the globally defined vectors, which are isomorphic to the Lorentz group.
These vectors were dubbed ``super-rotations,'' and, analogously to the supertranslations, they are a kind of asymptotic angle-dependent rotations and Lorentz boosts.
To maintain the algebra structure of these asymptotic symmetries, the supertranslations must be correspondingly modified.
A second extended symmetry group, due to Campiglia and Laddha~\cite{Campiglia:2014yka,Campiglia:2015yka}, considers all the diffeomorphisms of the 2-sphere rather than those equal to the Lorentz transformations, but the supertranslations are the same as in the BMS group.
The 2-sphere diffeomorphisms are often referred to as super Lorentz transformations~\cite{Compere:2018ylh}.

Both the super-rotations and super Lorentz transformations have corresponding conserved charges.
The charges for both algebras have been called ``super angular momentum,'' but they have also been called simply super-rotation charges or super Lorentz charges, for the respective algebras.
We shall primarily focus on the generalized BMS algebra, and we shall refer to the charges associated with this algebra as the super angular momentum (and will call those associated with the super-rotations the ``super-rotation charges.'').
Note that we will call the split of the charges into their electric- and magnetic-parity parts by super CM and superspin, respectively, in analogy with the convention used initially in~\cite{Flanagan:2015pxa} for the super-rotation charges, and subsequently for the super angular momentum in~\cite{Nichols:2017rqr,Nichols:2018qac}.\footnote{This is a second discrepancy with the nomenclature used in~\cite{Compere:2019gft}.
There, what we call superspin is called super angular momentum, and what we call super angular momentum is called a super Lorentz charge.
Our usages of super center-of-mass are equivalent, however.}

The super-rotation charges have a similar form to the angular momenta, but a super-rotation vector field enters into the expression for the charge rather than a Lorentz vector field (see, e.g.,~\cite{Barnich:2011mi,Flanagan:2015pxa}).
The super Lorentz charges constructed defined in~\cite{Compere:2018ylh} also have a similar form to the angular momentum with the Lorentz vector field is replaced by a super Lorentz transformation, but they have an additional term linear in the shear tensor needed to satisfy a flux balance law~\cite{Compere:2018ylh}.
Given that there is a one-parameter family of angular momentum that satisfies a number of reasonable physical conditions, it is also natural to ask whether there is such a parametrization for the super angular momentum.
We investigate this issue as well by allowing for a two-parameter family of super angular momentum that generalizes the Hamiltonian definition of~\cite{Compere:2018ylh} in a way completely analogous to the two-parameter extension of the Wald-Zoupas angular momentum given in~\cite{Compere:2019gft}.
In this case, setting the parameters to be equal (thereby reducing it to a one-parameter family) does not seem to make the super Lorentz charges vanish.
This is consistent with a calculation performed by Comp\`ere and Long~\cite{Compere:2016hzt} for the Hamiltonian charges.
There is a choice of parameters that makes the super angular momentum vanish, but this choice does not correspond to the Hamiltonian definition of~\cite{Compere:2018ylh}.
Rather, this choice is the same as the one used in~\cite{Compere:2020lrt} to determine a representation of the extended BMS algebra in nonradiative regions of spacetime for the super Lorentz charges in terms of the standard Poisson bracket.
This also leads to the possibility that properties of the generalized BMS algebra and charges could provide a criteria to prefer a certain definition of the angular momentum
(though we will not discuss this possibility further in this paper; see instead~\cite{Compere:2021inq}).

We then compute the multipolar expansion of the difference of the two-parameter family of super angular momentum from the Hamiltonian super angular momentum of~\cite{Compere:2018ylh}.
This allows us to see that unlike the angular momentum, the change in the difference in the super angular momentum will be nonvanishing even in stationary regions.
As a concrete example, we estimate the value of the change in the difference of the super angular momentum for nonspinning, quasicircular BBH mergers.
The relative size of the net change in Hamiltonian value of the super angular momentum and the net change in the difference term is small for these BBH mergers (a roughly one-percent effect).
Although it is small, it can be resolved given the current accuracy of numerical relativity (NR) simulations.

\paragraph*{Overview}
The outline of the rest of this paper is as follows.
Section~\ref{sec:BondiSachs} is mostly a review in which we introduce Bondi coordinates, the metric in these coordinates, the evolution equations for the Bondi mass and angular-momentum aspects, the (extended) BMS symmetries of asymptotically flat spacetimes, and the expressions for the various definitions of angular momentum in Bondi coordinates. 
We end this section, however, by introducing the proposed two-parameter definition of the super angular momentum. 
In Sec.~\ref{sec:flat_calc}, we compute the (super) angular momentum in flat spacetime (where we show two of the parameters must be equal for the angular momentum to vanish).
In the next section, Sec.~\ref{sec:MultipoleExpand}, we perform a multipolar expansion of the (super) angular momentum that is valid for general asymptotically flat spacetimes. 
In Sec.~\ref{sec:PNandNR}, we estimate the effect that the remaining free parameter in the angular momentum and super angular momentum has on BBH mergers of different mass ratios.
We compute results in the post-Newtonian approximation and using NR surrogate waveforms.
We conclude in Sec.~\ref{sec:Conclusions}.
In Appendix~\ref{app:STF_spherical_harmonics_conversion}, we compare our multipolar expansion of the angular momentum with a related expansion performed in~\cite{Compere:2019gft}.
In this paper, we use geometric units $G=c=1$, and the conventions on the metric and curvature tensors in~\cite{Wald:1984rg}.

\section{Bondi-Sachs framework, symmetries, and charges}
\label{sec:BondiSachs}

In this section, we review aspects of the Bondi-Sachs framework including the metric, some components of Einstein's equations, the asymptotic symmetries, and the corresponding charges.
We then discuss different definitions of angular momentum and super angular momentum.

\subsection{Metric and Einstein's equations}
 
We will perform our calculations in Bondi coordinates~\cite{Bondi:1962px,Sachs:1962wk} $(u,r,\theta^A)$, where $A=1,2$, and we review the properties of these coordinates and the solutions of Einstein's equations below.
We will use the notation and conventions given in~\cite{Flanagan:2015pxa}.
The metric in these coordinates is written in the form
\begin{align} \label{eq:BondiMetric}
ds^2 = & -U e^{2\beta} du^2-2e^{2\beta} du dr\nonumber\\
& +r^2 \gamma_{AB} (d\theta^A -U^A du)(d\theta^B - U^B du) \, 
\end{align}
where the functions and tensors $U$, $\beta$, $\gamma^{AB}$, and $U^A$ depend on all four Bondi coordinates $(u,r,\theta^A)$. 
The metric by construction satisfies the Bondi gauge conditions $g_{rr}=0$ and $g_{rA}=0$; Bondi coordinates also are defined such that $\det(\gamma_{AB})=\gamma(\theta^A)$ is independent of $u$ and $r$.
Some important properties of these coordinates are that $u$ is a retarded time variable (i.e., $u=$const.~are null hypersurfaces), $r$ is an areal radius, and $\theta^A$ (with $A=1,2$) are coordinates on 2-spheres of constant $r$ and $u$.

Near future null infinity (i.e., where $r$ is large), the metric functions $U$, $\beta$, $\gamma_{AB}$, and $U^A$ can be expanded as series in $1/r$. 
Asymptotically flat solutions postulate a given form of the expansion of these Bondi metric functions. 
For the tensor $\gamma_{AB}$ the conditions of asymptotic flatness generally impose
\begin{equation}
    \label{eq:gamma_AB_expansion}
    \gamma_{AB} = h_{AB}+\frac{1}{r}C_{AB}+O(r^{-2}) \, ,
\end{equation}
where $h_{AB}(\theta^C)$ is the metric on the unit 2-sphere, $C_{AB}$ is a function of $(u,\theta^A)$, and the determinant condition of Bondi gauge implies that $C_{AB} h^{AB} = 0$.
The remaining functions $U$, $\beta$, and $U^A$ are assumed to have the following limits as $r$ approaches infinity\footnote{Although we consider generalized BMS charges in this paper, we still impose the standard boundary conditions of asymptotic flatness and assume $h_{AB}$ is the round 2-sphere metric with constant Ricci scalar curvature and $U$ approaches unity as $r$ approaches infinity.
We restrict to these conditions, because we consider binary-black-hole mergers in this paper.
These are asymptotically flat solutions that remain in a fixed super Lorentz frame, and we then restrict to the trivial super Lorentz rest frame of the system.
Even with this restriction on the set of super Lorentz frames, the super angular momentum is nontrivial for these spacetimes.
If one considers a space of solutions that are  super Lorentz transformed from the boundary conditions given here, then one would need to consider the more general set of boundary conditions given, e.g., in~\cite{Compere:2018ylh}}
\begin{equation}
    \label{eq:hypersurfaceBCs}
    \lim_{r\rightarrow\infty} \beta = \lim_{r\rightarrow\infty} U^A = 0 \, , \qquad \lim_{r\rightarrow\infty} U = 1 \, .
\end{equation}

We will now specify to vacuum spacetimes to discuss Einstein's equations, for simplicity. 
The $ru$, $rA$, and trace of the $AB$ components of Einstein's equations take the form of hypersurface equations that can be solved on surfaces of constant $u$ by integrating radially outward.
The form of these equations is summarized in the review~\cite{Madler:2016}, for example.
The results of substituting Eq.~\eqref{eq:gamma_AB_expansion} into these hypersurface equations, radially integrating, and applying the boundary conditions in Eq.~\eqref{eq:hypersurfaceBCs} gives the following solutions for the remaining functions $U$, $\beta$, and $U^A$:
\begin{subequations} \label{eq:UbetaUA}
\begin{align}
\beta={}& -\frac{1}{32r^2}C_{AB}C^{AB}+O(r^{-3}) \, ,\\
U={}& 1-\frac{2m}{r}+O(r^{-2}) \, ,\\
\label{eq:U_A_expansion}
U^A={}& -\frac{1}{2r^2}D_B C^{AB}+\frac{1}{r^3}\Big[-\frac{2}{3}N^A +\frac{1}{16}D^A(C_{BC}C^{BC})\nonumber\\
&+\frac{1}{2} C^{AB}D^CC_{BC}\Big]+O(r^{-4}) \, .
\end{align}
\end{subequations}
We have introduced a number of new pieces of notation in the above equation, which we will now explain:
First, the function $m(u,\theta^A)$ is the Bondi mass aspect and $N^A(u,\theta^B)$ is the angular momentum aspect.
They are related to ``functions of integration'' that arise from integrating the hypersurface equations radially.
Second, in the above equation, we have raised and lowered indices of tensors and vectors on the 2-sphere using the metric $h^{AB}$ (respectively $h_{AB}$).
Third, we have defined the derivative operator $D_A$ as the torsion-free, metric-compatible derivative associated with the metric $h_{AB}$.

The evolution equation for $\gamma_{AB}$, when expanded to leading order in $1/r$, shows that the $u$ derivative of $C_{AB}$ is unconstrained by Einstein's equations and is defined to be the Bondi news tensor $N_{AB}=\partial_u C_{AB}$. 
The leading-order parts of the $uu$ and $uA$ components of Einstein equations are the conservation equations, which look like 
evolution equations for the Bondi mass aspect $m$ and the angular momentum aspect $N_A$ at fixed radii:
\begin{subequations} \label{eq:dotNAdotm}
\begin{align}
\label{eq:mevolve}
\dot{m}={}& -\frac{1}{8}N_{AB}N^{AB}+\frac{1}{4}D_AD_BN^{AB}\\
\label{eq:NAevolve}
\dot{N}_A={}&  D_A m + \frac{1}{4} D_B D_A D_C C^{BC} -\frac{1}{4} D_B D^B D^C C_{CA} \nonumber\\
&+ \frac{1}{4} D_B (N^{BC} C_{CA}) + \frac{1}{2} D_B N^{BC} C_{CA}
\end{align}
\end{subequations}
These equations are important for establishing flux balance laws for the charges conjugate to the asymptotic symmetries that form the BMS group and its extensions; we turn to the subject of these symmetries in the next subsection.

\subsection{Asymptotic symmetries}

The Bondi-Metzner-Sachs (BMS) group~\cite{Bondi:1962px,Sachs:1962zza} can be obtained from set of transformations that preserve the Bondi gauge conditions of the metric~\eqref{eq:BondiMetric} and the asymptotic form of the functions that appear in the metric [Eqs.~\eqref{eq:gamma_AB_expansion} and~\eqref{eq:UbetaUA}]. 
The BMS group is the semidirect product of the infinite-dimensional abelian group of supertranslations with a six-dimensional group of conformal transformations of the 2-sphere (which is isomorphic to the proper, isochronous Lorentz group). 
The four spacetime translations are a subgroup of the supertranslation group.
More recent generalizations of the BMS algebra take two forms.
(i) The first is the extended BMS algebra proposed by Barnich and Troessaert~\cite{Barnich:2009se,Barnich:2010eb,Barnich:2011mi} (see also~\cite{Banks:2003vp}).
In this proposal, all conformal Killing vectors of the 2-sphere are added to the algebra, including those with complex-analytic singularities on the 2-sphere. 
These additional symmetry vector fields were dubbed super-rotations, and the vectors that are isomorphic to the Lorentz transformations are a subalgebra of the super-rotations.
The supertranslations also are extended to include functions that are not necessarily smooth.
(ii) The second proposal has been called the generalized BMS algebra, and is due to Campiglia and Laddha~\cite{Campiglia:2014yka,Campiglia:2015yka}.
Here all smooth diffeomorphisms of the 2-sphere are considered instead of those equivalent to the Lorentz transformations, but the supertranslations are the same as in the original BMS group (though it is no longer possible to identify a preferred spacetime translation subgroup~\cite{Flanagan:2019vbl}).

The BMS symmetries and their generalizations are described by infinitesimal vector fields $\vec\xi$ that formally are defined at future null infinity, the null boundary of an asymptotically flat spacetime in the covariant conformal approach of Penrose~\cite{Penrose:1962ij,Penrose:1965am}. 
The form of the vector fields at future null infinity can be written in Bondi coordinates by restricting the vector fields that preserve the Bondi gauge conditions and the fall off rates of the metric to the tangent space of surfaces of constant $r$, and then taking the limit as $r$ goes to infinity. 
In this limit, the vector fields for the BMS group and its extensions all take the same form; they are parameterized by a scalar function $T(\theta^A)$ and a vector on the 2-sphere $Y^A(\theta^B)$:
\begin{equation}
    \vec\xi=\left[T(\theta^A)+\frac{1}{2}uD_AY^A(\theta^B) \right] \vec\partial_u + Y^A(\theta^B) \vec\partial_A
\end{equation}
The function $T(\theta^A)$ parametrizes the supertranslations in the BMS algebra and its generalizations (for the standard and generalized BMS algebras, it is assumed to be a smooth function, whereas for the extended BMS algebra, it can have complex analytic singular points). 
The vector field $Y^A(\theta^B)$ is a conformal Killing vector on the 2-sphere for the standard and extended BMS algebras (it is spanned by a six-parameter basis for the standard BMS algebra, or an infinite dimensional basis for the extended BMS algebra), or a smooth vector field for the generalized BMS group.

The symmetries at future null infinity can also be extended into the interior of the spacetime at large, but finite $r$ by requiring that the diffeomorphisms generated by these vector fields preserve the Bondi gauge conditions and the asymptotic fall-off conditions imposed on the metric.
Under these transformations, the functions $C_{AB}$, $N_{AB}$, $m$, and $N_A$ transform in a nontrivial way.
For the discussion that follows, we will only need the transformation law for $C_{AB}$, and we denote this transformation by $C_{AB} \rightarrow C_{AB} + \delta_\xi C_{AB}$, which was derived, e.g., in~\cite{Barnich:2010eb}.
It is convenient to first define a quantity
\begin{equation}
    f = T + \frac u2 D_A Y^A \, ,
\end{equation}
which appears in $\delta_\xi C_{AB}$ as follows:
\begin{align} \label{eq:deltaCABxi}
    \delta_\xi C_{AB} = {} & f N_{AB} - (2D_A D_B - h_{AB} D^2)f \nonumber \\
    & + \mathcal L_Y C_{AB} - \frac 12 D_C Y^C C_{AB} \, .
\end{align}
This transformation of $C_{AB}$ is useful for defining fluxes of conserved quantities associated with the BMS symmetries, which we will discuss in the next subsection.
Before we do so, it is useful to introduce a decomposition of the tensor $C_{AB}$ into its electric and magnetic (parity) parts as follows:
\begin{equation} \label{eq:CABdecomp}
   C_{AB}= \left(D_A D_B -\frac{1}{2}h_{AB}D^2\right)\Phi + \epsilon_{C(A}D_{B)} D^C \Psi \, .
\end{equation}
The scalars $\Phi$ and $\Psi$ are both smooth functions of the coordinates $(u,\theta^A)$.
From the transformation of $C_{AB}$ in Eq.~\eqref{eq:deltaCABxi}, it follows that a supertranslation affects the electric part of $C_{AB}$, but leaves the magnetic part invariant.
This property of the shear has been understood for quite some time (see, e.g.,~\cite{Newman:1966ub}).

\subsection{Fluxes and charges}

There are a few different prescriptions used to define the charges and the fluxes of charges that are associated with BMS symmetries.
We will describe here the procedure of Wald and Zoupas~\cite{Wald:1999wa}, in which the charges and fluxes are computed using a generalization of Noether's theorem that allows for the charges to change from emitted fluxes of gravitational waves and other matter fields.
We denote the charges by $Q_\xi[\mathcal C]$, where the charges depend linearly upon a BMS vector field $\vec \xi$ and are defined on a cross section of null infinity $\mathcal C$ (in Bondi coordinates, a surface of constant $u$ at fixed $r$ in the limit of $r\rightarrow\infty$).
We call the flux $\mathcal F_\xi[\Delta \mathscr I]$.
Like the charge, it has a linear dependence on a BMS vector field $\vec\xi$, but the flux depends on a region of null infinity $\Delta\mathscr I$ between two cuts (in Bondi coordinates, the region between two surfaces of constant $u$ at fixed $r$ in the limit of $r\rightarrow\infty$).
The flux balance law for the charges requires that
\begin{equation} \label{eq:FluxBalance}
    Q_\xi[\mathcal C_2] - Q_\xi[\mathcal C_1] = \mathcal F_\xi[\Delta\mathscr I] \, .
\end{equation}
The explicit expression for the flux has a simple form in Bondi coordinates in vacuum (see, e.g.,~\cite{Flanagan:2015pxa})
\begin{equation} \label{eq:Flux}
    F_\xi[\Delta\mathscr I] = - \frac 1{32\pi} \int_{\Delta\mathscr I} du \, d^2\Omega \, N^{AB} \delta_\xi C_{AB} \, ,
\end{equation}
where $\delta_\xi C_{AB}$ is given in Eq.~\eqref{eq:deltaCABxi} and $d^2\Omega$ is the area element on the 2-sphere cuts of constant $u$.
Using Eq.~\eqref{eq:deltaCABxi} and the conservation equations for the Bondi mass and angular momentum aspects in Eq.~\eqref{eq:dotNAdotm}, it is possible to show that the charge is given by
\begin{align} \label{eq:xiCharge}
Q_\xi = {} & \frac{1}{8\pi} \int_{\mathcal C} d^2\Omega \, \bigg\{2 T m + Y^A \bigg[ N_A - u  D_A m \nonumber\\
& -\frac{1}{16} D_A (C_{BC}C^{BC}) -\frac{1}{4} C_{AB} D_C C^{BC}\bigg] \bigg\}
\end{align}
(again, see, e.g.,~\cite{Flanagan:2015pxa}).
We dropped the dependence of the charge on the cut $\mathcal C$ to simplify the notation, and because it is made explicit in the domain of the integral on the right-hand side of the equation.

When the vector field $\vec \xi$ has $Y^A=0$ and $T \neq 0$, then it is a supertranslation, and the corresponding charge is the supermomentum.
The other case, a vector field with $Y^A \neq 0$ and $T = 0$, has as its corresponding charge the angular momentum, when $Y^A$ is equivalent to a Lorentz transformation for the standard BMS group.
The angular momentum is often split into its intrinsic (or spin) and center-of-mass (CM) parts, which correspond to the rotation and boost symmetries in the Lorentz group, respectively.
It was observed in~\cite{Flanagan:2015pxa} that the charge in Eq.~\eqref{eq:xiCharge} does not satisfy the flux balance law~\eqref{eq:FluxBalance} for the extended or generalized BMS vector fields.
A charge that does satisfy a flux balance for the super Lorentz charges was determined in~\cite{Compere:2018ylh}.
It is the same as that in Eq.~\eqref{eq:xiCharge}, up to the addition of two new terms linear in the tensor $C_{AB}$, and it is given below:\footnote{The flux for which this charge satisfies the flux balance law differs from Eq.~\eqref{eq:Flux}. 
It is necessary to add a term of the form 
\begin{equation}
\frac 1{64\pi} \int_{\Delta\mathscr I} du \, d^2\Omega \, u (D^2 D^B N_{AB} - D_B D_A D_C N^{BC} ) 
\end{equation}
to the right-hand side of Eq.~\eqref{eq:Flux} to restore the balance law with the definition of the charge in Eq.~\eqref{eq:xiCharge2} (see~\cite{Compere:2019gft} for further details).}
\begin{align} \label{eq:xiCharge2}
Q_\xi = {} & \frac{1}{8\pi} \int_{\mathcal C} d^2\Omega \, \bigg\{2 T m + Y^A \bigg[ N_A - u  D_A m \nonumber\\
& -\frac{1}{16} D_A (C_{BC}C^{BC}) -\frac{1}{4} C_{AB} D_C C^{BC} \nonumber \\
& + \frac u8 (D^2 D^B C_{AB} - D_B D_A D_C C^{BC} )\bigg] \bigg\} \, .
\end{align}
Note that the integral of the two additional terms in the final line Eq.~\eqref{eq:xiCharge2} can be shown to vanish for the $Y^A$ corresponding to Lorentz vector fields; note also that these two terms in the integrand are proportional to a differential operator acting on the magnetic part $\Psi$ of the shear in Eq.~\eqref{eq:CABdecomp} (see, e.g.,~\cite{Flanagan:2015pxa}).
The super angular momentum in~\eqref{eq:xiCharge2} can be divided into a magnetic-parity part called superspin and an electric-parity part called super center-of-mass, in analogy to the standard angular momentum.
In the next subsection, we focus on the angular momentum and discuss a subtlety in its definition.

\subsection{Definitions of angular momentum and their properties}

As discussed in the introduction, the angular momentum computed by Wald and Zoupas is not the only notion of the angular momentum of an isolated system that is commonly used.
While a number of the different angular momenta are equivalent, not all the definitions agree.
First, for convenience, let us specialize the general BMS charges in Eq.~\eqref{eq:xiCharge2} to a vector field $\vec\xi$ with $T=0$ and $Y^A$ being a generator of Lorentz transformations:
\begin{align} \label{eq:QYcharge}
Q_Y = &\frac{1}{8\pi}\int_{\mathcal C} d^2\Omega \, Y^A \left[ N_A - u  D_A m -\frac{1}{16} D_A (C_{BC}C^{BC})\right.\nonumber\\
&\left.-\frac{1}{4} C_{AB} D_C C^{BC}\right] \, .
\end{align}
We used the notation $Q_Y$ rather than $Q_\xi$ to emphasize that it depends only on $Y^A$. 
It has been shown in~\cite{Wald:1999wa} that the flux of this angular momentum agrees with that of Ashtekar and Streubel~\cite{Ashtekar:1981bq} and the charge defined by Dray and Streubel~\cite{Dray:1984rfa} (which came from twistorial definitions of the angular momentum~\cite{Penrose:1982wp}).
The Landau-Lifshitz definition of angular momentum in~\cite{RevModPhys.52.299} (which is restricted to the center-of-mass frame of the source and averaged over a few wavelengths of the emitted gravitational waves) also agrees with the flux of the angular momentum charge in Eq.~\eqref{eq:QYcharge}, when the expression is restricted to this context~\cite{Nichols:2017rqr}.

There are a few notable examples of definitions of angular momentum that differ from the one in Eq.~\eqref{eq:QYcharge}, a fact that was recently pointed out in a paper by Comp\`ere \textit{et al.} in~\cite{Compere:2019gft}.
First, in the context of conservation laws of gravitational scattering, a definition of an angular momentum involving just the mass and angular momentum aspects and the vector field on the 2-sphere, $Y^A$, was used in~\cite{Pasterski:2015tva,Hawking:2016sgy} to define the (super) angular momentum: i.e.,
\begin{align} \label{eq:QNonly}
Q_Y^{(0)} = \frac{1}{8\pi}\int_{\mathcal C} d^2\Omega Y^A (N_A - u  D_A m)\, .
\end{align}
Also recently, a more general definition of the Landau-Lifshitz angular momentum was proposed by by Bonga and Poisson~\cite{Bonga:2018gzr}, who no longer required that the result be defined in the CM frame or by averaging over a few wavelengths of the gravitational waves.
They specialized to the intrinsic (as opposed to CM) angular momentum, which they defined by using a collection of vector fields on the 2-sphere, $Y^A_i = \epsilon^{AD}\partial_D n_i$.
Here $n_i$ is a unit vector normal to the 2-sphere in quasi-Cartesian coordinates constructed from the spatial Bondi coordinates $(r,\theta^A)$, and $\epsilon^{AD}$ is the Levi-Civita tensor on the unit 2-sphere.
After converting their definition of the intrinsic angular momentum into our notation, their result can be written as
\begin{equation} \label{eq:LLJi}
J_i= \frac{1}{8\pi} \! \int_{\mathcal C} \! \! d^2\Omega \epsilon^{AD}\partial_D n_i \! \left[ N_A - u  D_A m -\frac{3}{4} C_{AB} D_C C^{BC}\right]  .
\end{equation}
There is a definition of the CM part of the angular momentum in the Landau-Lifshitz formalism from Blanchet and Faye~\cite{Blanchet:2018yqa}, but it was shown in~\cite{Compere:2019gft} that it cannot easily be written in terms of the 2-sphere-covariant Bondi-metric functions.
As we discuss further below, the three definitions of the angular momentum in Eqs.~\eqref{eq:QYcharge}--\eqref{eq:LLJi} all vanish in flat spacetime, give the same angular momentum of a Kerr black hole  and satisfy flux balance laws; they thus appear to be equally viable definitions of the angular momentum of an isolated source.

Given that the angular momenta in Eqs.~\eqref{eq:QYcharge}--\eqref{eq:LLJi} differ in the factors in front of the two terms quadratic in $C_{AB}$ in Eq.~\eqref{eq:QYcharge}, Comp\`ere \textit{et al.}~\cite{Compere:2019gft} observed that a two-parameter family of charges could be defined by allowing the coefficients in front of these terms to be arbitrary real numbers. 
When the coefficients are restricted to specific values, the two-parameter family of charges reduces to one of the specific definitions in Eqs.~\eqref{eq:QYcharge}--\eqref{eq:LLJi}.
Thus, the two-parameter family of angular momentum of Comp\`ere \textit{et al.}~\cite{Compere:2019gft} is given by
\begin{align}
\label{eq:AngMomAlphaBeta}
Q_Y^{(\alpha,\beta)} = {} &\frac{1}{8\pi}\int_{\mathcal C} d^2\Omega Y^A \left[ N_A - u  D_A m -\frac{\alpha}{4} C_{AB} D_C C^{BC}\right.\nonumber\\
&\left.-\frac{\beta}{16} D_A (C_{BC}C^{BC})\right],
\end{align}
where $\alpha$ and $\beta$ are real constants.\footnote{The terms $D_A (C_{BC}C^{BC})$ and $C_{AB} D_C C^{BC}$ form a kind of basis of vectors constructed from contractions of $C_{AB}$ and $D_A C_{BC}$, in the sense that other possible contractions can be rewritten in terms of these two quantities~\cite{Compere:2019gft}.}
The Wald-Zoupas angular-momentum corresponds to the case $\alpha=\beta=1$; the angular momentum in Eq.~\eqref{eq:QNonly} corresponds to $\alpha=\beta=0$; and the intrinsic angular momentum in Eq.~\eqref{eq:LLJi} corresponds to $\alpha=3$ (and $\beta$ can take on any real value, because it does not contribute to the intrinsic part). 
For all values of $\alpha$ and $\beta$, the angular momentum in Eq.~\eqref{eq:AngMomAlphaBeta} satisfies flux balance laws, but it is not immediately apparent that they will vanish in flat spacetime.
In the next section, we will derive the conditions under which the angular momentum in Eq.~\eqref{eq:AngMomAlphaBeta} vanishes in flat spacetime.

\subsection{Definitions of super angular momentum}

The charge in Eq.~\eqref{eq:AngMomAlphaBeta} was defined specifically for the angular momentum. 
There are also differing definitions of the super angular momentum, however, because several of the definitions of the super angular momentum were defined through promoting the vector field $Y^A$ that enters into the charge from a Lorentz vector field to a super Lorentz vector.
The definition in Eq.~\eqref{eq:QNonly} was also used for a super-rotation charge (where $Y^A$ is a super-rotation vector field, for example), and this definition differs from that in Eq.~\eqref{eq:QYcharge}.
The main difference between the two charges is are the terms quadratic in the shear tensor.
It thus seems reasonable to define a two-parameter family of charges that satisfy a flux balance law by generalizing Eq.~\eqref{eq:xiCharge2} (when $T=0$) to include real coefficients $\alpha$ and $\beta$ in front of the terms quadratic in $C_{AB}$.
Thus, we will also consider a two-parameter family of super angular momentum defined by
\begin{align} \label{eq:QYalphabetaCharge}
Q_Y^{(\alpha,\beta)}= {} & \frac{1}{8\pi} \int_{\mathcal C} d^2\Omega \, Y^A \bigg[ N_A - u  D_A m \nonumber\\
& + \frac u8 (D^2 D^B C_{AB} - D_B D_A D_C C^{BC} ) \nonumber \\
& -\frac{\alpha}{4} C_{AB} D_C C^{BC} -\frac{\beta}{16} D_A (C_{BC}C^{BC}) \bigg] \, .
\end{align}
We will investigate the properties of this charge in flat spacetime next.

\section{(Super) angular momentum in flat spacetime}
\label{sec:flat_calc}

While the focus in this section will be determining the values of the coefficients $\alpha$ and $\beta$ for which the angular momentum vanishes in flat spacetime, much of the calculation holds for any smooth vector field on the 2-sphere $Y^A$, and thus applies to the super angular momentum of the generalized BMS algebra.\footnote{Note however that if $Y^A$ is a super-rotation vector field of the extended BMS algebra, then the singular points of the vector fields make integration by parts on the 2-sphere more challenging. 
Although the 2-sphere is a compact manifold without boundary, when integrating by parts one must carefully analyze the contributions that come from boundary-like terms at the singular points of the super-rotation vectors, which can contribute to the integral (see, e.g.,~\cite{Compere:2016jwb} for further details).}
In the derivation that follows, it is structured so that the first part applies to smooth generalized BMS vectors $Y^A$, and the next part is specified to $Y^A$ that generate Lorentz transformations.
Note that a similar calculation was performed by Comp\`ere and Long in~\cite{Compere:2016jwb} for the Wald-Zoupas charges (i.e., $\alpha = \beta = 1$).

In flat spacetime, there is no radiation, and the news tensor vanishes~\cite{Geroch:1977jn}.
In this case, the Bondi mass aspect and the Bondi angular momentum are also proportional to components of the vacuum Riemann tensor (see, e.g.,~\cite{Flanagan:2015pxa}) and thus they must also vanish. 
From Eq.~\eqref{eq:NAevolve}, one can then also show that $\Psi$, the scalar that parametrizes the magnetic part of $C_{AB}$ must also vanish.
Because $C_{AB}$ is electric type, then by performing a supertranslation it follows from Eq.~\eqref{eq:deltaCABxi} that it is possible to choose a frame in which the tensor $C_{AB}$ vanishes (note that from the transformation properties of $m$ and $N_A$ given in, e.g.,~\cite{Flanagan:2015pxa}, the mass and angular momentum aspects will remain zero under this transformation).
We will not work in the frame where $C_{AB}$ vanishes, but rather we will choose a frame where it has a nonzero electric part.
Thus, the values of the relevant functions needed to compute the super angular momentum in Eq.~\eqref{eq:QYalphabetaCharge} are given by
\begin{subequations}
\label{eq:canonical_frame}
\begin{align}
\label{eq:canonical_frame_mass}
m={}&0 \, , \\
\label{eq:canonical_frame_angmom}
N_{A}={}&0 \, ,\\
\label{eq:canonical_frame_shear}
C_{AB}={}& \left(D_A D_B -\frac{1}{2}h_{AB} D^2 \right)\Phi \, .
\end{align}
\end{subequations}
In flat spacetime, therefore, the additional terms in the second line of Eq.~\eqref{eq:QYalphabetaCharge} do not contribute, and the super angular momentum is given by
\begin{align}
\label{eq:charge_definition}
Q_Y^{(\alpha,\beta)} = -\frac{1}{128\pi}\int_{\mathcal C} d^2\Omega & \left[4 \alpha Y^A C_{AB} D_C C^{BC} \right. \nonumber\\
&\left. +\beta Y^A D_A  (C_{BC} C^{BC})\right] \, .
\end{align}

We will now substitute in the expression in Eq.~\eqref{eq:canonical_frame_shear} for $C_{AB}$ in Eq.~\eqref{eq:charge_definition} in several places, and begin simplifying the expression.
Because we are assuming $Y^{A}$ is a smooth vector on the 2-sphere and $\Phi$ is a smooth function, we can integrate the first term by parts and drop the terms involving divergences of vector fields on the 2-sphere. 
For the second term, we use the fact that the covariant derivative acting on the shear tensor in Eq.~\eqref{eq:canonical_frame_shear} is given by
\begin{equation}
    D^B C_{AB} = D^B D_A D_B \Phi - \frac{1}{2} D_A D^2 \Phi \, .
\end{equation}
We can then use the definition of the Riemann tensor (associated with the derivative operator $D_A$) to commute the first two covariant derivatives in the first term.
We find that it can be written as
\begin{equation}
\label{eq:DBCAB}
    D^B C_{AB} = D_A D^2 \Phi + R_{AB} D^B \Phi - \frac{1}{2} D_A D^2 \Phi \, ,
\end{equation}
where $R_{AB}$ is the Ricci tensor on the 2-sphere.
Assuming that the metric is that of a round 2-sphere, then the scalar curvature of the sphere is given by $R=2$, the Ricci tensor is $R_{AB}=h_{AB}$, and the Riemann tensor can be written as
\begin{equation}
    R_{ABCD} = h_{AC} h_{BD} - h_{AD} h_{BC} \, .
\end{equation}
This implies that $D^B C_{AB}$ simplifies to
\begin{equation}
    \label{eq:derivative_shear}
    D^B C_{AB} = \frac{1}{2} D_A (D^2+2) \Phi \, .
\end{equation}
Next, substituting Eqs.~\eqref{eq:derivative_shear} and~\eqref{eq:canonical_frame_shear} into Eq.~\eqref{eq:charge_definition}, we can write the charge in terms of $Y^A$, $\Phi$, and derivative operators $D_A$ (though we leave one term involving $C_{AB}$). 
If we integrate by parts once more for both the terms proportional to $\alpha$ and $\beta$, we find the super angular momentum is given by
\begin{align} \label{eq:QYmid}
& Q_Y^{(\alpha,\beta)} = \nonumber \\ 
&\frac{1}{128\pi}\int_{\mathcal C} d^2\Omega \left\{\beta D_A Y^A  [D_B D_C\Phi D^B D^C\Phi - \frac{1}{2}(D^2 \Phi)^2] \right.\nonumber\\
& \left. + 2 \alpha \left[D^B Y^A C_{AB} + \frac 12 Y^A D_A (D^2+2)\Phi \right] (D^2+2)\Phi \right\}.
\end{align}
While for each $\Phi$ and $Y^A$ there should exist a choice of $\alpha$ and $\beta$ that makes $Q_Y$ vanish, a choice of $\alpha$ and $\beta$ that makes the super angular momentum vanish for all $\Phi$ and $Y^A$ in flat spacetime is $\alpha=\beta=0$.
However, it is not necessarily clear that one should require that the super angular momentum should vanish, as Comp\`ere and collaborators have argued that the super angular momentum can be used to distinguish vacuum states that differ by a supertranslation~\cite{Compere:2016jwb,Compere:2018ylh}.
We thus only identify $\alpha=\beta=0$ as a choice that makes the super angular momentum vanish in flat spacetime, but do not require the charge to satisfy this property.

\paragraph*{Angular momentum}
We do require that the charge $Q_Y$ vanish for vectors $Y^A$ that generate Lorentz transformations.
We now continue our simplification of Eq.~\eqref{eq:QYmid} by using the fact that $Y^A$ is a conformal Killing vector on the 2-sphere; i.e., it satisfies the conformal Killing equation
\begin{equation} \label{eq:CKV_YA}
    2D_{(A}Y_{B)}-D_C Y^C h_{AB}=0 \, .
\end{equation}
Because $C_{AB}$ is symmetric and trace free, then $C_{AB} D^B Y^A$ involves only the symmetric-trace-free part of $D^B Y^A$.
By the conformal Killing equation~\eqref{eq:CKV_YA}, however, $D^B Y^A$ is proportional to $h_{AB}$, so $C_{AB} D^B Y^A$ vanishes.
After performing a large number of integration by parts (so as to write the expression mostly in terms of squares of $\Phi$ and its derivatives) and using the following identity
\begin{equation}
    D^2 D^C \Phi = D^C D^2 \Phi + D^C \Phi \, ,
\end{equation}
we find that the angular momentum can be written as
\begin{align}
\label{eq:charge}
Q_Y^{(\alpha,\beta)} = \frac{1}{256\pi} \int_{\mathcal C} d^2\Omega & \left\{ (D_A Y^A) \big[(\beta - \alpha)(D^2 \Phi )^2 - 4\alpha \Phi^2 \right.\nonumber\\
&+2(2\alpha - \beta)D_C \Phi D^C \Phi\big] \nonumber \\
& - 2D^2 (D_A Y^A) \big[\alpha \Phi^2 - \beta D_C\Phi D^C \Phi \big] \nonumber\\
&- \left. 2\beta D_B D_C D_A Y^A D^B\Phi D^C\Phi \right\}.
\end{align}
Conformal Killing vectors also satisfy the property that 
\begin{equation}
\label{eq:property1ofkillingvector}
    (D^2 + 2 )(D_A Y^A) = 0 \, ,
\end{equation}
which leads to the cancellation of some terms proportional to $\alpha$ in Eq.~\eqref{eq:charge}.
The globally defined conformal Killing vectors (the vector fields $Y^A$ that can be written as a superposition of the six $l=1$ vector spherical harmonics on the 2-sphere) satisfy the additional property
\begin{equation}
\label{eq:property2ofkillingvector}
    D_B D_C D_A Y^A = - h_{BC} D_A Y^A \, .
\end{equation}
After using the property in Eq.~\eqref{eq:property2ofkillingvector} in Eq~\eqref{eq:charge}, we see that the angular momentum in flat spacetime can be written as 
\begin{equation}
Q_Y^{(\alpha,\beta)} = \frac{1}{256\pi}(\beta - \alpha) \int_{\mathcal C} d^2\Omega D_A Y^A [ (D^2 \Phi )^2- 4D_C \Phi D^C \Phi ] \, .
\end{equation}
The intrinsic angular momentum (i.e., the charge $Q_Y^{(\alpha,\beta)}$ for vectors $Y^A$ with $D_A Y^A = 0$) vanishes for all values of $\alpha$ and $\beta$.
For the center-of-mass angular momentum (i.e., the charge with $Y^A$ that has nonvanishing $D_A Y^A$), the charge will typically be nonvanishing unless $\alpha=\beta$.
Having the physical requirement that the angular momentum should vanish in flat spacetime thus reduces the two-parameter family of charges to a one-parameter family given by $\alpha$.
We will typically work with this reduced one-parameter family in the rest of the paper, unless we note otherwise.

We conclude this section with an important note.
Because our expressions for the mass and angular momentum aspects vanish in flat spacetime, our calculations in this section apply to the $\alpha$- and $\beta$-dependent terms in any region of spacetime, where the tensor $C_{AB}$ can be written in terms of the electric part as in Eq.~\eqref{eq:canonical_frame_shear}.
While this section is then nominally about flat spacetime, the results in this part directly imply that the different definitions of angular momentum that vanish in flat spacetime will all agree in any region of spacetime with electric shear (stationary or radiative). 
In particular, the result that the angular momentum vanishes when $\alpha=\beta$ in flat spacetime means that in stationary regions, the angular momenta for any real value of $\alpha$ will be equivalent.
Requiring the angular momentum takes on a particular value in a particular stationary solution cannot be used to restrict this remaining parameter $\alpha$.
For the angular momentum, we will then focus on the differences that arise in radiative regions with magnetic-parity shear.
For the super angular momentum, which only manifestly vanishes when $\alpha = \beta = 0$, there can be differences in its value for distinct $\alpha$ and $\beta$ values for the same spacetime, as we also illustrate in more detail below.

\section{Multipolar expansion of the (super) angular momentum}
\label{sec:MultipoleExpand}

We will first summarize our conventions for the spherical harmonics that we use in our multipolar expansion. 
We will then perform multipolar expansions of the super angular momentum, which we will subsequently specialize to the standard angular momentum.

Because the multipolar expansion of Hamiltonian charges and fluxes had been computed previously (see, e.g.,~\cite{Nichols:2017rqr,Nichols:2018qac,Compere:2019gft}), we will focus on the difference of the two-parameter family of charges from the charge defined in~\cite{Compere:2018ylh}.
Thus, for a vector field $Y^A$ we will write
\begin{subequations}
\begin{equation}
    Q_Y^{(\alpha,\beta)} =  Q_Y^{(\alpha=1,\beta=1)} + (\alpha-1) \delta Q_Y^{(\alpha=1)} + (\beta-1) \delta Q_Y^{(\beta=1)} \, ,
\end{equation}
where $Q_Y^{(\alpha=1,\beta=1)}$ is the charge with $\alpha=\beta=1$ and $\delta Q_Y^{(\alpha=1)}$ and $\delta Q_Y^{(\beta=1)}$ are defined by
\begin{align}
\label{eq:deltaQYalpha}
    \delta Q_Y^{(\alpha=1)} = & -\frac{1}{32\pi} \int_{\mathcal C} d^2\Omega \, Y^A C_{AB} D_C C^{BC} \, , \\
    \label{eq:deltaQYbeta}
    \delta Q_Y^{(\beta=1)} = & -\frac{1}{128\pi} \int_{\mathcal C} d^2\Omega \, Y^A D_A (C_{BC}C^{BC}) \, .
\end{align}
\end{subequations}
In the special case of angular momentum, we will also use the notation $\delta J_Y^{(\alpha=1)}$ and $\delta k_Y^{(\alpha=1)}$ (and similarly for the $\beta$ term) for the difference in the intrinsic and CM angular momentum, respectively, associated with a vector $Y^A$ (which is a rotation or Lorentz boost, respectively).
A similar calculation was performed in~\cite{Compere:2019gft}; however, here we also compute the $\alpha$-dependent term in the CM angular momentum, and we write the result in terms of the multipole moments $U_{lm}$ and $V_{lm}$ (defined below) rather than the rank-$l$ symmetric-trace-free (STF) tensors $U_L$ and $V_L$ (discussed in Appendix~\ref{app:STF_spherical_harmonics_conversion}).
The moments $U_{lm}$ and $V_{lm}$ are somewhat easier to relate to the moments of the gravitational-wave strain $h_{lm}$ that can be obtained from numerical-relativity simulations or surrogate models fit to simulations (the latter of which we will use later in Sec.~\ref{sec:PNandNR}). 

In the cases where we restrict to $\alpha=\beta$ (so that the angular momentum vanishes in flat spacetime), then we will use the notation 
\begin{subequations}
\begin{equation} \label{eq:QYalpha_equals_beta}
    Q_Y^{(\alpha=\beta)} =  Q_Y^{(\alpha=\beta=1)} + (\alpha-1) \delta Q_Y^{(\alpha=\beta=1)}  \, ,
\end{equation}
where $Q_Y^{(\alpha=\beta=1)} = Q_Y^{(\alpha=1,\beta=1)}$ is the charge with $\alpha=\beta=1$ and $\delta Q_Y^{(\alpha=\beta=1)}$ is defined by
\begin{align} \label{eq:deltaQYalpha_equals_beta}
    \delta Q_Y^{(\alpha=\beta=1)} = -\frac{1}{128\pi} \int_{\mathcal C} d^2\Omega \, Y^A & [4 C_{AB} D_C C^{BC} \nonumber \\
    & + D_A (C_{BC}C^{BC})] \, .
\end{align}
\end{subequations}
We will similarly use the notation $\delta J_Y^{(\alpha=\beta=1)}$ and $\delta k_Y^{(\alpha=\beta=1)}$ for the intrinsic and CM angular momentum, respectively, when $Y^A$ is a rotation or Lorentz boost (also respectively).

\subsection{Spherical harmonics and multipolar expansion of the gravitational-wave data}

In addition to the scalar spherical harmonics (with the usual Condon-Shortly phase convention), $Y_{lm}(\theta,\phi)$, we will use vector and tensor harmonics on the unit 2-sphere, which we define as in~\cite{Nichols:2017rqr}.
The vector harmonics are given by 
\begin{subequations}
\begin{align} 
\label{eq:TAelmdef}
    T^{A}_{(e),lm} = {} & \frac{1}{\sqrt{l(l+1)}}D^{A}Y_{lm} \, ,\\
    \label{eq:TAblmdef}
    T^{A}_{(b),lm} = {} & \frac{1}{\sqrt{l(l+1)}}\epsilon^{AB} D_{B}Y_{lm} \, ,
\end{align}
\end{subequations}
which are nonzero for $l\geq 1$ and the tensor harmonics
\begin{subequations}
\begin{align}
  \label{eq:electric_tensor_harmonics}  
  T_{AB}^{(e),lm} = {} & \frac 12 \sqrt{\frac{2(l-2)!}{(l+2)!}}\left(2D_A D_B - h_{AB} D^2\right) Y_{lm} \, ,\\
  \label{eq:TABblmdef}
  T_{AB}^{(b),lm} = {} & \sqrt{\frac{2(l-2)!}{(l+2)!}} \epsilon_{C(A}D_{B)}D^C Y_{lm} \, ,
\end{align}
\end{subequations}
which are nonzero for $l\geq 2$. 

We use these harmonics to expand the shear tensor as
\begin{equation}
\label{eq:multipole_expansion_shear}
C^{AB}=\sum_{l,m} (U_{lm} T^{AB}_{(e),lm}+V_{lm} T^{AB}_{(b),lm}) \, .
\end{equation}
Because the shear is real, the coefficients in this expansion obey the properties
\begin{equation}
\label{eq:conj_multipole_moments}
U_{l,-m}=(-1)^m \bar{U}_{lm} \, , \quad V_{l,-m}=(-1)^m \bar{V}_{lm} \, ,
\end{equation}
where the overline means to take the complex conjugate.
By using Eqs.~\eqref{eq:TAelmdef}--\eqref{eq:TABblmdef} and~\eqref{eq:DBCAB}, we can write the covariant derivative of the shear tensor in terms of vector harmonics as follows:
\begin{align}
\label{eq:shear_covariant_derivative}
D_{C} C^{BC}=\sum_{l,m} \sqrt{\frac{(l-1)(l+2)}{2}} (U_{lm} T^{B}_{(e),lm}-V_{lm} T^{B}_{(b),lm}). 
\end{align}

The vector and tensor harmonics are related to spin-weighted spherical harmonics ${}_sY_{lm}$ of spin weight $s=\pm 1$ and $s=\pm 2$, respectively, and a complex null dual vector on the 2-sphere 
\begin{equation} \label{eq:mdyad}
m^A \partial_A = \frac 1{\sqrt{2} }(\partial_\theta + i \csc\theta \partial_\phi)  \, .
\end{equation} 
and its complex conjugate $\bar m^A$. 
The relationships for the vector harmonics are
\begin{subequations}
\begin{align}
    \label{eq:electric_vector_harmonic}
    T_A^{(e),lm} = {} & \frac{1}{\sqrt{2}}\left(\leftidx{_{-1}}Y_{lm} m_A - \leftidx{_{1}}Y_{lm} \bar{m}_A\right) \, , \\
    \label{eq:magnetic_vector_harmonic}
    T_A^{(b),lm} = {} & \frac{i}{\sqrt{2}}\left(\leftidx{_{-1}}Y_{lm} m_A + \leftidx{_{1}}Y_{lm} \bar{m}_A\right) \, ,
\end{align}
\end{subequations}
and for the tensor harmonics are 
\begin{subequations}
\begin{align}
    T_{AB}^{(e),lm} = {} & \frac{1}{\sqrt{2}}\left( \leftidx{_{-2}}Y_{lm}m_Am_B + \leftidx_2Y_{lm}\bar{m}_A \bar{m}_B \right) \, ,\\
    \label{eq:TABblm2}
    T_{AB}^{(b),lm} = {} & -\frac{i}{\sqrt{2}}\left( \leftidx{_{-2}}Y_{lm}m_Am_B - \leftidx_2Y_{lm}\bar{m}_A \bar{m}_B \right) \, .
\end{align}
\end{subequations}
The spin-weighted spherical harmonics satisfy the well-known complex-conjugate property ${}_s\bar Y_{lm} = (-1)^{s+m}{}_{-s} Y_{l-m}$.

The charges are quadratic in $C_{AB}$ and involve a vector field $Y^A$, and we will expand all three quantities in terms of spin-weighted spherical harmonics using Eqs.~\eqref{eq:TAelmdef}--\eqref{eq:TABblm2}.
When evaluating the charges, we will frequently encounter integrals of three spin-weighted spherical harmonics over $S^2$.
We use the notation of~\cite{Nichols:2017rqr} to describe these integrals, which we denote by
\begin{align} \label{eq:CldefIntegral}
& C_l(s',l',m';s'',l'',m'') \equiv \nonumber\\
& \int d^2\Omega \, (_{s'+s''}\bar{Y}_{lm'+m''})(_{s'}{Y}_{l'm'})(_{s''}{Y}_{l''m''}) \, .
\end{align}
The complex-conjugated spherical harmonic $_{s'+s''}\bar{Y}_{lm'+m''}$ has spin-weight $s=s'+s''$ and azimuthal number $m=m'+m''$, because for all other values of $s$ and $m$, the integral vanishes.
It can be shown that the coefficients $C_l(s',l',m';s'',l'',m'')$ can be written in terms of Clebsch-Gordon coefficients $\langle l',m';l'',m''|l,m'+m'' \rangle$ as follows:
\begin{align} \label{eq:CldefClebschGordon}
C_l(s',l',m';s'',l'',m'')= (-1)^{l+l'+l''} \sqrt{\frac{(2l'+1)(2l''+1)}{4 \pi (2l+1)}} \nonumber\\
\times \left< l',s';l'',s''|l,s'+s'' \right> \left< l',m';l'',m''|l,m'+m'' \right> \, .
\end{align}
The coefficients are also nonvanishing only when the $l$ index is in the range $\{\max(|l'-l''|,|m'+m''|,|s'+s'' |),...,l'+l''-1,l'+l''\}$.
There are two additional useful identities under sign flips of the spin weight and azimuthal numbers that we will need in the discussion below
\begin{subequations}
\label{eq:Clsyms}
\begin{align}
    C_l(s',l',m';s'',l'',m'') = & (-1)^{l+l'+l''}\nonumber\\
    &\times C_l(-s',l',m';-s'',l'',m''),\\
    C_l(s',l',m';s'',l'',m'') = & (-1)^{l+l'+l''}\nonumber\\
    &\times C_l(s',l',-m';s'',l'',-m'').
\end{align}
\end{subequations}
We can now turn to the evaluation of the terms $\delta Q_Y^{(\alpha=1)}$ and $\delta Q_Y^{(\beta=1)}$ in a few specific cases of interest next.

\subsection{Multipolar expansion of the super angular momentum}
\label{sec:extended_BMS_charge}

In this part, we will compute the multipolar expansion of the $\alpha$ and $\beta$ ``difference terms'' in Eqs.~\eqref{eq:deltaQYalpha} and~\eqref{eq:deltaQYbeta} from the super angular momentum of~\cite{Compere:2018ylh}.
We will consider two types of vector fields $Y^A$ to compute the charges: namely, the electric- and magnetic-parity vectors harmonics defined in Eqs.~\eqref{eq:TAelmdef} and~\eqref{eq:TAblmdef}.
We will thus denote these terms by $\delta Q^{(\alpha=1)}_{(e),lm}$ and $\delta Q^{(\alpha=1)}_{(b),lm}$, respectively, for Eq.~\eqref{eq:deltaQYalpha} and $\delta Q^{(\beta=1)}_{(e),lm}$ and $\delta Q^{(\beta=1)}_{(b),lm}$, respectively, for Eq.~\eqref{eq:deltaQYbeta}.
The results here hold for both the standard BMS charges (CM and intrinsic angular momentum) and the generalized BMS charges (super angular momentum).
There are a number of additional simplifications that occur for the intrinsic and CM angular momentum, and we will therefore treat these simpler cases separately afterwards.

In this calculation, we will not require initially that the two parameters $\alpha$ and $\beta$ be equal, because this choice was made to require that the standard (rather than the super) angular momentum vanishes in flat spacetimes. 
For the super angular momentum, the choice of $\alpha=\beta$ does not guarantee that these charges vanish in flat spacetimes, and it is not agreed upon universally that these charges should vanish in flat spacetime (see, e.g.,~\cite{Compere:2016jwb}). 

Before we begin the calculations, note that because $D^A T^{(b),lm}_A = 0$, then by performing an integration by parts of Eq.~\eqref{eq:deltaQYbeta}, one can show that 
\begin{equation}
    \delta Q^{(\beta=1)}_{(b),lm} = 0 \, ;
\end{equation}
we will thus focus on the three quantities $\delta Q^{(\alpha=1)}_{(e),lm}$, $\delta Q^{(\alpha=1)}_{(b),lm}$, and $\delta Q^{(\beta=1)}_{(e),lm}$.
The calculation of these three quantities is quite similar, so we will describe in detail the procedure for just $\delta Q^{(\alpha=1)}_{(e),lm}$ (and the other two quantities can be determined through a nearly identical calculation).

Starting from Eq.~\eqref{eq:deltaQYalpha}, we then substitute in the multipolar expansion of $C_{AB}$ and $D_A C^{AB}$ given in Eqs.~\eqref{eq:multipole_expansion_shear} and~\eqref{eq:shear_covariant_derivative} and the vector spherical harmonic in Eq.~\eqref{eq:TAelmdef}.
We then use the relationships between the vector and tensor spherical harmonics and the spin-weighted spherical harmonics in Eqs.~\eqref{eq:electric_vector_harmonic}--\eqref{eq:TABblm2} to write $\delta Q^{(\alpha=1)}_{(e),lm}$ in terms of the multipole moments $U_{lm}$ and $V_{lm}$ as well as the integrals of three spin-weighted spherical harmonics in Eq.~\eqref{eq:CldefIntegral}.
We then make use of the identities for the coefficients $C_l(s',l',m';s'',l'',m'')$ in Eq.~\eqref{eq:Clsyms} and the complex conjugate properties of $U_{lm}$ and $V_{lm}$ in Eq.~\eqref{eq:conj_multipole_moments} to simplify the expression.
It is useful to make the definitions (similar to those in~\cite{Nichols:2018qac})
\begin{subequations}
\begin{align}
    s^{l,(\pm)}_{l';l''} = {} & 1 \pm (-1)^{l+l'+l''} \, ,\\
    f^l_{l',m';l'',m''} = {} & \sqrt{(l'+2)(l'-1)}C_l(-1,l',m';2,l'',m'') \, ,\\
    g^l_{l',m';l'',m''} = {} & \sqrt{l(l+1)}C_l(-2,l',m';2,l'',m'') \, .
\end{align}
\end{subequations}
The result can then be written as is
\begin{subequations}
\begin{align} \label{eq:deltaQalphae}
    \delta Q^{(\alpha=1)}_{(e),lm} = -\frac 1{128\pi} & \sum_{l',m';l'',m''} f^l_{l',m';l'',m''} \nonumber \\
    & \times  [ s^{l,(+)}_{l';l''} (U_{l'm'}U_{l''m''} + V_{l'm'}V_{l''m''}) \nonumber \\
    & +i  s^{l,(-)}_{l';l''} (U_{l'm'}V_{l''m''} - V_{l'm'}U_{l''m''}) ] \, ,
\end{align}
where the indices on the charges should be integers in the ranges $l\geq 1$ and $-l\leq m \leq l$, and where the sums run over integers in the ranges $l'\geq 2$, $-l'\leq m' \leq l'$, $l''\geq 2$, and $-l''\leq m'' \leq l''$
This gives the $\alpha$-dependent difference from the super-CM charge of~\cite{Compere:2018ylh}.
A similar calculation shows that the $\alpha$-dependent correction to the superspin can be written as  
\begin{align} \label{eq:deltaQalphab}
    \delta Q^{(\alpha=1)}_{(b),lm} = \frac i{128\pi} & \sum_{l',m';l'',m''} f^l_{l',m';l'',m''} \nonumber \\
    & \times  [ s^{l,(-)}_{l';l''} (U_{l'm'}U_{l''m''} + V_{l'm'}V_{l''m''}) \nonumber \\
    & +i  s^{l,(+)}_{l';l''} (U_{l'm'}V_{l''m''} - V_{l'm'}U_{l''m''}) ] \, .
\end{align}
Finally, the $\beta$-dependent correction to the super-CM charge is given by
\begin{align} \label{eq:deltaQbetae}
    \delta Q^{(\beta=1)}_{(e),lm} = -\frac 1{256\pi} & \sum_{l',m';l'',m''} g^l_{l',m';l'',m''} \nonumber \\
    & \times  [ s^{l,(+)}_{l';l''} (U_{l'm'}U_{l''m''} + V_{l'm'}V_{l''m''}) \nonumber \\
    & +i  s^{l,(-)}_{l';l''} (U_{l'm'}V_{l''m''} - V_{l'm'}U_{l''m''}) ] \, .
\end{align}
\end{subequations}
The values of $l$, $l'$, $l''$, $m$, $m'$, and $m''$ in Eqs.~\eqref{eq:deltaQalphab} and~\eqref{eq:deltaQbetae} are the same as in Eq.~\eqref{eq:deltaQalphae}. 
From these difference terms and the super-CM and superspin charges with $\alpha=1$ and $\beta=1$ (i.e., $Q^{(\alpha=1,\beta=1)}_{(e),lm}$ and $Q^{(\alpha=1,\beta=1)}_{(b),lm}$) one can then construct the full $\alpha$ and $\beta$ dependent super CM and superspin (i.e., $Q^{(\alpha,\beta)}_{(e),lm}$ and $Q^{(\alpha,\beta)}_{(b),lm}$).

Although we do not require that the superspin and super CM vanish in flat spacetime, it is still useful to write down the expressions for the $\alpha$- and $\beta$-dependent difference terms in this case: namely, the quantities $\delta Q^{(\alpha=\beta=1)}_{(e),lm}$ and $\delta Q^{(\alpha=\beta=1)}_{(b),lm}$.
It is then straightforward to specialize our previous results to find that
\begin{subequations}
\begin{align}
\label{eq:superCMalphaEQbeta}
    \delta Q^{(\alpha=\beta=1)}_{(e),lm} = -\frac 1{256\pi} & \sum_{l',m';l'',m''} \! \! \!
    (2f^l_{l',m';l'',m''}+g^l_{l',m';l'',m''}) \nonumber \\
    & \times  [ s^{l,(+)}_{l';l''} (U_{l'm'}U_{l''m''} + V_{l'm'}V_{l''m''}) \nonumber \\
    & +i  s^{l,(-)}_{l';l''} (U_{l'm'}V_{l''m''} - V_{l'm'}U_{l''m''}) ] \, .
\end{align}
The superspin is the same, because the term $\delta Q^{(\beta=1)}_{(b),lm}$ vanishes: i.e.,
\begin{equation}
\label{eq:superspinalphaEQbeta}
    \delta Q^{(\alpha=\beta=1)}_{(b),lm} = 
    \delta Q^{(\alpha=1)}_{(b),lm} \, .
\end{equation}
\end{subequations}
In the next subsections, we will further specialize Eqs.~\eqref{eq:superCMalphaEQbeta} and~\eqref{eq:superspinalphaEQbeta} to $l=1$ spherical harmonics to compute the CM and intrinsic angular momentum.

\subsection{Multipolar expansion of the intrinsic angular momentum}

We begin by simplifying the expression in Eq.~\eqref{eq:deltaQalphab} in the case where $l=1$ (which corresponds to the correction to the intrinsic angular momentum).
When $l=1$, the coefficients $f^1_{l',m';l'',m''}$ are nonvanishing for $l''=l'$ or $l''=l'\pm 1$.
Thus, the coefficient $s^{1,(-)}_{l';l''}$ is nonvanishing only when $l''=l'$ and the coefficient $s^{1,(+)}_{l';l''}$ is nonvanishing for $l''=l'\pm 1$.
Because the index $m$ satisfies $m=0$ or $m=\pm 1$, then for the first set of terms in Eq.~\eqref{eq:deltaQalphab} proportional to $s^{1,(-)}_{l';l''}$ the nonzero terms in the double sum will be one of the terms of the form $f^1_{l',m';l',-m'}$ or $f^1_{l',m';l',-m'\pm1}$.
Given the complex-conjugate relationships for the $U_{lm}$ and $V_{lm}$ moments in Eq.~\eqref{eq:conj_multipole_moments} and the symmetries of the coefficients $C_1(-1,l',m';2,-l',m'')$ under the change of sign of $m'$ in Eq.~\eqref{eq:Clsyms}, then one can show that the terms proportional to $s^{1,(-)}_{l';l''}$ vanish.
The difference term from the Wald-Zoupas angular momentum is then given by
\begin{align}
\delta J_{1,m}^{(\alpha=1)} \equiv \delta Q_{(b),1,m}^{(\alpha=1)} = {} & \frac{1}{128\pi}\sum\limits_{l',m',l'',m''} s^{1,(+)}_{l';l''} f^1_{l',m';l',m''} \nonumber\\ 
&\times ({U}_{l'm'}V_{l'',m''}-{V}_{l'm'}U_{l'',m''}) \, .
\end{align}
Note that although we left the expression as a double sum over $l'$ and $l''$, the $l''$ sum is restricted to $l''=l'-1$ or $l''=l'+1$; similarly, the $m''$ sum is restricted to the values $m''=m-m'$, where $m=0$ or $m=\pm 1$.
If we evaluate the coefficients $f^1_{l',m';l'\pm 1,-m'}$, $f^1_{l',m';l'\pm 1,-m'-1}$, and $f^1_{l',m';l'\pm 1,-m'+1}$ in the sum using the expression in Eq.~\eqref{eq:CldefClebschGordon}, then the expressions can be simplified to square roots of rational functions in these cases.
We follow~\cite{Nichols:2018qac} and define coefficients $a_l$, $b_{lm}^{(\pm)}$, $c_{lm}$ and $d_{lm}^{(\pm)}$ by
\begin{subequations}
\label{eq:angmomcoeffs}
\begin{align} 
    a_l= {} & \sqrt{\frac{(l-1)(l+3)}{(2l+1)(2l+3)}} \, , \\
    b_{lm}^{(\pm)}= {} & \sqrt{(l\pm m+1)(l\pm m+2)} \, , \\
    c_{lm}= {} & \sqrt{(l-m+1)(l+m+1)} \, , \\
    d_{lm}^{(\pm)}= {} & \sqrt{(l\pm m +1)(l\mp m)}
\end{align}
\end{subequations}
(though we do not use $d_{lm}^{(\pm)}$ until the next subsection).
In terms of these quantities, and after relabelling $l'$ with $l$ and $m'$ with $m$ in the sum, we can write the difference term from the Wald-Zoupas angular momentum as
\begin{subequations}
\label{eq:intrinsic_angular_momentum}
\begin{align}
\label{eq:intrinsic_angular_momentum_(1,0)moment} \delta J_{1,0}^{(\alpha=1)} = \frac{1}{16}\sqrt{\frac{3}{2\pi}} & \sum\limits_{l\geq 2,m}\frac{a_l c_{lm}}{l+1}
 \nonumber\\ 
&\times (\bar{U}_{lm}V_{l+1,m}-\bar{V}_{lm}U_{l+1,m}) \, ,\\
\label{eq:intrinsic_angular_momentum_(1,pm1)moment} \delta J_{1,\pm1}^{(\alpha=1)} = \frac{1}{32}\sqrt{\frac{3}{\pi}} & \sum\limits_{l\geq 2,m}\frac{a_l b_{lm}^{(\pm)}}{l+1}
 \nonumber\\ 
& \times (\bar{U}_{lm}V_{l+1,m\pm1}-\bar{V}_{lm}U_{l+1,m\pm1}) \, .
\end{align}
\end{subequations}
The calculation to arrive at these simplified expressions requires some relabelling of indices in the sum so that only terms with $l+1$ appear rather than $l-1$.

A similar calculation was performed in~\cite{Compere:2019gft} using STF $l$-index tensors rather than expanding $C_{AB}$ in the harmonics in Eq.~\eqref{eq:multipole_expansion_shear}. 
The two formalisms can be related, and we compared the result of the difference term in~\cite{Compere:2019gft} for the intrinsic angular momentum to our expressions in Eqs.~\eqref{eq:intrinsic_angular_momentum_(1,0)moment} and~\eqref{eq:intrinsic_angular_momentum_(1,pm1)moment}.
We found that our result differs from Eq.~(4.16) of~\cite{Compere:2019gft} by an additional factor of $1/(l+1)$, and we could not identify from where this discrepancy was arising. 
We give a detailed calculation of this comparison in Appendix~\ref{app:STF_spherical_harmonics_conversion}.
Given our results in the next subsection, we believe our result to be correct, so we suspect that the error lies in the conversion between the two formalisms.

\subsection{Multipolar expansion of the center-of-mass angular momentum}

We now derive a similar expression for the difference terms from the Wald-Zoupas center-of-mass angular momentum when expanded in terms of the the mass and current multipole moments of $C_{AB}$ in Eq.~\eqref{eq:multipole_expansion_shear}.
We first give a result for general real coefficients $\alpha$ and $\beta$, and we then specify to  the $\alpha=\beta$ choice.
The calculation is quite similar to that in the previous subsection for the intrinsic angular momentum. 
When the expression in Eq.~\eqref{eq:deltaQalphae} is restricted to $l=1$, then there is again a similar cancellation of the terms proportional to $s^{1,(-)}_{l';l''}$ leaving just the terms proportional to $s^{1,(+)}_{l';l''}$.
Again, because the allowed values of $l''$ are given by $l''=l'\pm 1$, the coefficients $f^1_{l',m';l'',m''}$ simplify to square roots of rational functions.
The $\alpha$-dependent difference terms are then given by
\begin{subequations}
\label{eq:k10and11alpha}
\begin{align}
\delta Q_{(e),1,0}^{(\alpha=1)} \equiv \delta k_{1,0}^{(\alpha=1)} = {} & \frac{1}{16}\sqrt{\frac{3}{2\pi}} \sum\limits_{l\geq 2,m}\frac{a_l c_{lm}}{l+1} \nonumber \\
&\times (\bar{U}_{lm}U_{l+1,m}+\bar{V}_{lm}V_{l+1,m}) \, , \\
\delta Q_{(e),1,\pm 1}^{(\alpha=1)} \equiv \delta k_{1,\pm 1}^{(\alpha=1)} = {} & \frac{1}{32}\sqrt{\frac{3}{\pi}} \sum\limits_{l\geq 2,m}\frac{a_l b_{lm}^{(\pm)}}{l+1} \nonumber\\ 
&\times (\bar{U}_{lm}U_{l+1,m\pm1}+\bar{V}_{lm}V_{l+1,m\pm1}) \, ,
\end{align}
\end{subequations}
for the $m=0$ and $m=\pm 1$ modes, respectively.

For the $\beta$-dependent difference term in Eq.~\eqref{eq:deltaQbetae}, it is no longer the case that the $s^{1,(-)}_{l';l''}$ terms vanish. 
However, because the coefficients $g^1_{l',m';l'',m''}$ also have the property that they vanish except when $l''=l'$ or $l''=l'\pm 1$ and when $m''=m-m'$ for $m=0$ or $m=\pm 1$, then the coefficients can similarly be evaluated in terms of rational functions and their square roots.
The result of this calculation is as follows:
\begin{subequations}
\label{eq:CM_angular_momentum_w/out_parameters_condition}
\begin{align}
\delta Q_{(e),1,0}^{(\beta=1)} \equiv \delta k_{1,0}^{(\beta=1)} = {} & -\frac{1}{16}\sqrt{\frac{3}{2\pi}}\sum\limits_{l\geq 2,m}\frac{1}{l+1}\nonumber\\
&\times \bigg[a_l c_{lm}(\bar{U}_{lm}U_{l+1,m}+\bar{V}_{lm}V_{l+1,m}) \nonumber\\
&-\frac{2im}{l}\bar{U}_{lm} V_{lm} \bigg] \, , \\
\delta Q_{(e),1,\pm 1}^{(\beta=1)} \equiv \delta k_{1,\pm 1}^{ (\beta=1)} = {} & - \frac{1}{32}\sqrt{\frac{3}{\pi}}\sum\limits_{l\geq 2,m}\frac{1}{l+1}  \nonumber\\ 
&\times \bigg[a_l b_{lm}^{(\pm)}(\bar{U}_{lm}U_{l+1,m\pm1} \nonumber \\ 
& +\bar{V}_{lm}V_{l+1,m\pm1}) \nonumber\\
&\pm \frac{2i}{l}d_{lm}^{(\pm)}\bar{U}_{lm} V_{l,m\pm1} \bigg] \, .
\end{align}
\end{subequations}
The coefficients $d_{lm}^{(\pm)}$ are defined in Eq.~\eqref{eq:angmomcoeffs}.

A significant simplification occurs when the two parameters are equal; only the terms involving products of $U_{lm}$ and $V_{lm}$ moments remain.
We find that the result is given by
\begin{subequations}
\label{eq:CM_angular_momentum}
\begin{align}
\delta Q_{(e),1,0}^{(\alpha=\beta=1)} \equiv \delta k_{1,0}^{(\alpha=\beta=1)} = {} & \frac{i}{8}\sqrt{\frac{3}{2\pi}}\sum\limits_{l\geq 2,m}\frac{m}{l(l+1)}
\bar{U}_{lm} V_{lm} \, ,\\
\delta Q_{(e),1,\pm 1}^{(\alpha=\beta=1)} \equiv \delta k_{1,\pm 1 }^{(\alpha=\beta=1)} = {} & \mp  \frac{i}{16}\sqrt{\frac{3}{\pi}} \sum\limits_{l\geq 2,m}\frac{d_{lm}^{(\pm)}}{l(l+1)}\bar{U}_{lm} V_{l,m\pm1} \, .
\end{align}
\end{subequations}
This result is consistent with our calculation in flat spacetime in Sec.~\ref{sec:flat_calc}.
In that section, we showed that when $\alpha=\beta$, the angular momentum should vanish in flat spacetime.
Because the tensor $C_{AB}$ can be decomposed using just electric-type tensor harmonics (i.e., the $U_{lm}$ modes can be nonvanishing but all $V_{lm}$ modes must vanish), then the multipolar expansion should not involve products of $U_{lm}$ moments with other $U_{lm}$ moments, because these terms would be nonvanishing in flat spacetime.

Our result for the $\beta$-dependent term in Eqs.~\eqref{eq:CM_angular_momentum_w/out_parameters_condition} agrees with Eq.~(4.17) of~\cite{Compere:2019gft} after performing the same conversion between their STF $l$-index tensors and our mass and current multipoles $U_{lm}$ and $V_{lm}$.
This comparison is given in detail in Appendix~\ref{app:STF_spherical_harmonics_conversion}.
The $\alpha$-dependent terms in Eq.~\eqref{eq:k10and11alpha} was not computed in~\cite{Compere:2019gft}. 
Note, however, that the coefficients in $\delta k_{1m}^{(\alpha=1)}$ in Eq.~\eqref{eq:k10and11alpha} that multiply the products of $U_{lm}$ and $V_{lm}$ moments are precisely the same ones that appear in Eq.~\eqref{eq:intrinsic_angular_momentum} for $\delta J_{1m}^{(\alpha=1)}$.
Since the coefficients are the same in Eqs.~\eqref{eq:intrinsic_angular_momentum} and~\eqref{eq:k10and11alpha}, and since these coefficients are needed to have the angular momentum vanish in flat spacetime, then this provides a consistency check on the result in Eq.~\eqref{eq:intrinsic_angular_momentum}.

Now that we have the multipolar expressions for the difference terms from the Wald-Zoupas definition of the angular momentum, it is possible to assess how large these terms are for different systems of interest.
We will focus on nonspinning compact binaries in the next section.

\section{\label{sec:PNandNR} Standard and super angular momentum for nonprecessing BBH mergers}

In this part, we compute the effect of the remaining free parameter $\alpha$ on the standard and super angular momentum from nonprecessing binary-black-hole mergers.
As discussed in the introduction, the value of the (super) angular momentum depends on a choice of Bondi frame.
For the explicit calculations using PN theory and NR surrogate models in this section, we will work in the canonical frame (e.g.,~\cite{Flanagan:2015pxa}) associated with the binary as $u\rightarrow -\infty$.
This frame is a type of asymptotic rest frame in which $C_{AB}=0$ and the system has vanishing mass dipole moment (i.e., a CM frame). 

For the difference of the angular momentum from the Wald-Zoupas values [i.e., Eqs.~\eqref{eq:intrinsic_angular_momentum} and~\eqref{eq:CM_angular_momentum}], this difference depends on products of both the $U_{lm}$ and the $V_{lm}$ modes. 
As we discuss in the first subsection in this part, the $U_{lm}$ modes can be nonvanishing after the passage of GWs for these BBH mergers, because of the GW memory effect.
The $V_{lm}$ modes vanish after the radiation passes for these BBH systems (see, e.g.,~\cite{Mitman:2020pbt}; thus, the difference terms in Eqs.~\eqref{eq:intrinsic_angular_momentum} and~\eqref{eq:CM_angular_momentum} will vanish after the passage of the GWs. 
This implies that the net change in the angular momentum between two nonradiative regions for these binaries will be the same.
Nevertheless, while the binary is emitting GWs, the instantaneous value of the angular momentum will differ from the Wald-Zoupas value.
We compute the size of this effect in the post-Newtonian (PN) approximation and using surrogate models fit to numerical-relativity (NR) simulations in the following subsections.

We then perform similar calculations involving the difference terms from the super angular momentum of~\cite{Compere:2018ylh}.
Because the super angular momentum terms in Eq.~\eqref{eq:deltaQalphae} involve products of $U_{lm}$ moments, then the super angular momentum can differ from the $\alpha=\beta=1$ values when there is the GW memory effect.
We thus estimate the magnitude of this difference in the PN approximation and from the dominant waveform modes from NR simulations.
As we will discuss further below, the effect is small compared to the change in the super angular momentum, but is within the numerical accuracy of the simulations.

Because we are interested in investigating the order-of-magnitudes of the effects rather than their precise values, we will generally work with the leading-order approximations to the results in this section, as we will describe in more detail in the relevant parts below.

\subsection{Computing the leading GW memory effect and spin memory effect} \label{subsec:MemoryModes}

In post-Newtonian theory, the GW memory effect and the spin memory effect have been computed, and the relevant results can be obtained from, e.g.,~\cite{Wiseman:1991ss} or~\cite{Nichols:2017rqr}, respectively.
For NR simulations, GW memory effects are not captured in most Cauchy simulations (see, e.g.,~\cite{Boyle:2019kee}) and the additional post-processing step of Cauchy-characteristic extraction~\cite{Bishop:1996gt} needs to be performed~\cite{Pollney:2010hs,Mitman:2020pbt} to get the memory effects directly from simulations.
However, by enforcing the flux balance laws in Eq.~\eqref{eq:FluxBalance}, one can determine constraints on the GW memory effects from waveforms that do not contain the memory (e.g.,~\cite{Nichols:2017rqr,Mitman:2020bjf}).
This approximate procedure is quite accurate~\cite{Mitman:2020pbt}.
We summarize our procedure for computing GW memory effects below.

\subsubsection{(Displacement) GW memory effect}

The GW memory effect can be computed by integrating the conservation equation for the Bondi mass aspect in Eq.~\eqref{eq:mevolve} with respect to $u$ [this equation contains equivalent information to the flux balance law~\eqref{eq:FluxBalance} for a basis of supertranslation vectors].
Integrating the term $D_A D_B N^{AB}$ in Eq.~\eqref{eq:mevolve} with respect to $u$ gives rise to a change in the shear, which we will denote by $D_A D_B \Delta C^{AB}$. 
This quantity $D_A D_B \Delta C^{AB}$ is constrained by changes in the mass aspect $\Delta m$ and the integrated flux of energy per solid angle (a term proportional to $\int du N_{AB} N^{AB}$; see, e.g.,~\cite{Flanagan:2015pxa} and references therein for further discussion).
This equation constrains only the electric part of $\Delta C_{AB}$, and for this reason it is convenient to write the memory using a single scalar function $\Delta \Phi$ as
\begin{equation}
    \label{eq:change_in_shear}
    \Delta C_{AB} = \left(D_A D_B - \frac 12 h_{AB} D^2\right) \Delta \Phi \, .
\end{equation}
It is then useful to expand $\Delta \Phi$ in scalar spherical harmonics $Y_{lm}$.
Once this is done, when the the operator $ (2D_A D_B - h_{AB}D^2)$ acts on these scalar harmonics, Eq.~\eqref{eq:change_in_shear} can be written in terms of the electric-parity tensor harmonics in Eq.~\eqref{eq:electric_tensor_harmonics} as
\begin{align}
\label{eq:shear_memory}
    \Delta C_{AB}=\sum_{l,m} \sqrt{\frac{(l+2)!}{2(l-2)!}} T_{AB}^{(e),lm} \Delta \Phi_{lm} \, .
\end{align}
By comparing Eq.~\eqref{eq:shear_memory} with Eq.~\eqref{eq:multipole_expansion_shear}, it is straightforward to see that the change in the $U_{lm}$ moments can be related to the $\Delta \Phi_{lm}$ modes via the relationship
\begin{equation} \label{eq:PhilmToUlm}
    \Delta U_{lm}=\sqrt{\frac{(l+2)!}{2(l-2)!}} \Delta \Phi_{lm} \, .
\end{equation}

Although both changes in the Bondi mass aspect and the flux of energy per solid angle produce GW memory effects, for nonprecessing BBH mergers, the flux term produces the much larger memory effect (i.e., the nonlinear memory is much larger than the linear memory; this is true in both the post-Newtonian approximation~\cite{Favata:2008yd} and in NR simulations~\cite{Mitman:2020pbt}).
For this reason, just the contributions from the nonlinear memory to $\Delta\Phi$ were computed in~\cite{Nichols:2017rqr}, and the result is given in terms of the mass and current multipole moments by
\begin{align} \label{eq:DeltaPhilm}
    \Delta \Phi_{lm}= & \frac 12 \frac{(l-2)!}{(l+2)!}\sum_{l',l'',m',m''} C_l(-2,l',m';2,l'',m'')\nonumber\\
    &\times \int_{-\infty}^{\infty} du \{2i s^{l,(-)}_{l';l''} \dot{U}_{l'm'}\dot{V}_{l''m''}\nonumber\\
    &+s^{l,(+)}_{l';l''}(\dot{U}_{l'm'}\dot{U}_{l'',m''}+\dot{V}_{l'm'}\dot{V}_{l'',m''})\} \, .
\end{align}
Both in the PN approximation and in NR simulations, the largest contribution to the GW memory effect from nonprecessing BBH mergers comes from terms involving products of $U_{22}$ and $U_{2-2}=\bar U_{22}$ modes in Eq.~\eqref{eq:DeltaPhilm}. 
The dominant memory effect produced by the $U_{22}$ mode appears in the $\Delta\Phi_{20}$ and $\Delta\Phi_{40}$ modes. 
Evaluating the appropriate coefficients in Eq.~\eqref{eq:DeltaPhilm} and using Eq.~\eqref{eq:PhilmToUlm}, we find that the leading GW memory effect in the mode $U_{20}$ is given by
\begin{subequations}
\label{eq:U20U40memory}
\begin{equation}
    \label{eq:U20memory}
    \Delta U_{20}=\frac{1}{42}\sqrt{\frac{15}{\pi}}\int_{-\infty}^{\infty} du |\dot{U}_{22}|^2 \, .
\end{equation}
The expression for the $U_{40}$ mode is given by
\begin{equation}
    \Delta U_{40}=\frac{1}{504\sqrt{5\pi}}\int_{-\infty}^{\infty} du |\dot{U}_{22}|^2 = \frac 1{60\sqrt 3} \Delta U_{20} \, .
\end{equation}
\end{subequations}
We will also consider quantities $U_{20}$ and $U_{40}$ which are obtained by integrating Eq.~\eqref{eq:U20U40memory} from $-\infty$ up to a finite retarded time $u$ rather than than taking the limit $u\rightarrow \infty$.

\subsubsection{GW modes that produce the spin-memory effect}

The other type of GW memory that we will need to consider in this paper is the GW spin memory effect.
Like the GW memory effect in the previous subsection, the spin memory effect can also be determined from the flux balance law in Eq.~\eqref{eq:FluxBalance}.
Unlike the displacement memory, the spin memory is constrained by changes in the super angular momentum (rather than the supermomentum) and the flux of angular momentum per solid angle (rather than the flux of energy per solid angle).
In addition, the spin memory effect appears in the magnetic-parity part of the retarded-time integral of the shear tensor, rather than the electric part of the change in the shear.
We will not need the spin memory itself, but we do need the GW modes that produce the spin memory effect.
Nevertheless, it is easiest to describe the calculation of these modes by summarizing the calculation of the spin memory. 
We thus begin by writing the shear tensor $C_{AB}$ as a sum of two terms of electric- and magnetic-parity parts
\begin{align}
    C_{AB}=\frac 12 \left(2D_A D_B - h_{AB} D^2\right) \Phi + \epsilon_{C(A}D_{B)}D^C \Psi \, ,
\end{align}
where $\Phi$ and $\Psi$ are smooth functions of the coordinates $(u,\theta^A)$. 
The spin memory is related to the retarded time integral of the function $\Psi$~\cite{Nichols:2017rqr}
\begin{align}
    \Delta \Sigma \equiv \int_{-\infty}^{+\infty} du \, \Psi \, .
\end{align}
The full multipolar expansion of the spin memory is a somewhat lengthy expression, so we do not reproduce it here (although it is given in~\cite{Nichols:2017rqr}).
Analogously to the displacement GW memory effect, there are two contributions to the spin memory effect from the linear and nonlinear terms.
However, the linear terms are smaller than the nonlinear terms for nonprecessing compact binaries (see, e.g.,~\cite{Mitman:2020pbt}), so we focus on just the nonlinear terms.
We will also give just the largest terms that are computed from the mode $U_{22}$ (which is the dominant term in the PN approximation, and also the most significant term in NR simulations).
The $U_{22}$ mode produces a spin memory effect that appears in the $u$ integral of the $l=3$, $m=0$ mode of the waveform; it was computed in~\cite{Nichols:2017rqr} to be
\begin{equation}
    \Delta \Sigma = \frac{1}{80\sqrt{7\pi}} Y_{30} \int du \Im(\bar{U}_{22}\dot{U}_{22}) \, .
\end{equation}
Acting on $\Delta \Sigma$ with the operator $\epsilon_{C(A}D_{B)}D^C$ gives the retarded-time integral of the magnetic-parity part of the shear tensor $C_{AB}$: 
\begin{equation}
\label{eq:magnetic_part_CAB_spin_memory}
    \epsilon_{C(A}D_{B)}D^C \Delta \Sigma = \frac{1}{40} \sqrt{\frac{15}{7\pi}} T_{AB}^{(b),30} \int du \Im(\bar{U}_{22}\dot{U}_{22}) \, .
\end{equation}
By differentiating Eq.~\eqref{eq:magnetic_part_CAB_spin_memory} with respect to $u$, we can obtain the magnetic part of the shear that produces the spin memory effect.
Because Eq.~\eqref{eq:magnetic_part_CAB_spin_memory} is already expanded in magnetic-parity tensor harmonics, we can immediately determine that the relevant spin-memory mode is $V_{30}$, which is given by 
\begin{equation} \label{eq:V30spinmemory}
    V_{30} = \frac{1}{40} \sqrt{\frac{15}{7\pi}} \Im(\bar{U}_{22}\dot{U}_{22}) \, .
\end{equation}
We will use Eqs.~\eqref{eq:U20U40memory} and~\eqref{eq:V30spinmemory} to add in the contributions of the memory and spin memory effects that are not included in the NR surrogate waveform model that we use to compute the difference terms from the respective Hamiltonian definitions of~\cite{Wald:1999wa} and of~\cite{Compere:2018ylh} for the angular momentum and super angular momentum in the next subsections.

\subsection{Standard angular momentum}

We noted above that the different definitions of the angular momentum for nonprecessing BBH mergers will agree after the gravitational waves pass, but they will differ while these systems are radiating gravitational waves. 
We will calculate the size of this difference first in the post-Newtonian (PN) approximation and second in full general relativity using numerical-relativity waveforms from BBH mergers. 
The NR waveforms are usually given in terms of the multipole moments of the strain $h$, which is related to the tensor $C_{AB}$ by the relation
\begin{equation}
    h \equiv h_+ -i h_\times = \frac 1r C_{AB} \bar{m}^A\bar{m}^B \, .
\end{equation}
This expression defines the two polarizations $h_+$ and $h_\times$ and $\bar m^A$ is the complex conjugate of the dyad defined in Eq.~\eqref{eq:mdyad}.
The strain $h$ can be expanded in terms of spin-weighted spherical harmonics ${}_{-2}Y_{lm}$ as
\begin{equation}
    h =  \sum_{lm} h_{lm} \left(\leftidx{_{-2}}Y_{lm} \right)\, .
\end{equation}
It then follows that the moments $h_{lm}$ are related to $U_{lm}$ and $V_{lm}$ by
\begin{equation}
\label{eq:strain_radiative_multipole_moments_relation}
    h_{lm} = \frac{1}{r\sqrt{2}}\left(U_{lm}-iV_{lm}\right)
\end{equation}
(see, e.g.,~\cite{Nichols:2017rqr} and references therein).

Because of the symmetries of nonprecessing binaries, the relationship between the $h_{lm}$ mode and the $U_{lm}$ and $V_{lm}$ modes simplifies. 
Specifically, the mass multipole moments $U_{lm}$ are nonzero only when $l+m$ is even, and the current multipole moments $V_{lm}$ are nonzero only when $l+m$ is odd (see, e.g.,~\cite{Blanchet:2013haa}). 
Therefore, the mass and current multipole moments can be written in terms of the strain modes for these systems as
\begin{subequations}
\label{eq:htoUVlm}
\begin{align}
    U_{lm}= {} &r \sqrt{2}h_{lm} \, , \quad \, \mbox{for } l+m \mbox{ even} \, , \\
    V_{lm}= {} & ir \sqrt{2}h_{lm} \, , \quad \mbox{for } l+m \mbox{ odd} \, .
\end{align}
\end{subequations}
Note that our definition of the polarizations $h_+$ and $h_\times$ (and hence $h_{lm}$) have a relative minus sign to those in~\cite{Blanchet:2013haa}, though the $U_{lm}$ and $V_{lm}$ moments agree in sign.
Combining these properties of the $U_{lm}$ and $V_{lm}$ moments with the expressions for the difference terms in Eqs.~\eqref{eq:intrinsic_angular_momentum} and~\eqref{eq:CM_angular_momentum}, we find that multipole moments $\delta J_{1\pm1}^{(\alpha=\beta=1)}$ and $\delta k_{10}^{(\alpha=\beta=1)}$ vanish. 
Thus, we focus on the $\delta J_{10}^{(\alpha=\beta=1)}$ and $\delta k_{1\pm1}^{(\alpha=\beta=1)}$ modes below.

The waveforms from PN calculations and surrogate models from NR simulations contain a finite number of $(l,m)$ modes [in the PN context, the waveform has only been computed up to a finite PN order, whereas for surrogate models, the NR simulations extract only a subset of all $(l,m)$ modes, and the surrogate models only fit to a further subset of the extracted modes].
The number of modes that we use in the calculations of the quantities $\delta J_{10}^{(\alpha=\beta=1)}$ and $\delta k_{1\pm1}^{(\alpha=\beta=1)}$ will differ, but it is chosen such that we capture the leading nonvanishing effect in the PN approximation.
We will then use the same set of modes for the calculations with the NR surrogate waveform (absent any modes that the surrogate model does not contain).
As we will discuss in more detail below, we will use waveform modes that go up to 2.5PN orders above the leading part of the $U_{22}$ mode to compute $\delta J_{10}^{(\alpha=\beta=1)}$, whereas for $\delta k_{1\pm1}^{(\alpha=\beta=1)}$, we can capture the leading effect using just the leading $U_{22}$ mode and the $V_{21}$ mode. 
Thus, to compute $\delta J_{10}^{(\alpha=\beta=1)}$ we use the expression
\begin{align}
\label{eq:delta_J10_radiative_multipole_moments}
  \delta J_{10}^{(\alpha=\beta=1)} = {} \frac{1}{8}\sqrt{\frac{3}{2\pi}} & \Re\bigg[ \frac{a_2 c_{22}}{3} \bar{U}_{22}V_{32} + \frac{a_3 c_{33}}{4} \bar{U}_{33}V_{43}\nonumber\\
  &+ \frac{a_3 c_{31}}{4} \bar{U}_{31}V_{41}+ \frac{a_2 c_{20}}{6} \bar{U}_{20}V_{30} \nonumber\\
  & - \frac{a_2 c_{21}}{3} \bar{V}_{21}U_{31}
  -\frac{a_3 c_{32}}{4} \bar{V}_{32}U_{42} \nonumber\\
  & -\frac{a_4 c_{43}}{5} \bar{V}_{43}U_{53}-\frac{a_3 c_{30}}{8} \bar{V}_{30}U_{40}\bigg] \, ,
\end{align}
Note that the real part of the quantity in parentheses is being taken, which arises from using the complex-conjugate properties of the modes $U_{lm}$ and $V_{lm}$ in Eq.~\eqref{eq:conj_multipole_moments}.
For $\delta k_{1\pm1}^{(\alpha=\beta=1)}$, we use the expressions
\begin{subequations}
\label{eq:delta_k_radiative_multipole_moments}
\begin{align}
    \delta k_{11}^{(\alpha=\beta=1)} = \frac{i}{96} \sqrt{\frac{3}{\pi}} \left( d^{(+)}_{2-2} U_{22} \bar{V}_{21} - d^{(+)}_{20} \bar{U}_{20} V_{21}\right),\\
    \delta k_{1-1}^{(\alpha=\beta=1)} = \frac{i}{96} \sqrt{\frac{3}{\pi}} \left( d^{(-)}_{22} \bar{U}_{22} V_{21} - d^{(-)}_{20} U_{20} \bar{V}_{21}\right).
\end{align}
\end{subequations}
Here note that $\delta k_{11}^{(\alpha=\beta=1)} = - \delta \bar k_{1-1}^{(\alpha=\beta=1)}$, since $\delta k_{11}^{(\alpha=\beta=1)}$ and $\delta k_{1-1}^{(\alpha=\beta=1)}$ can both be related to the real difference terms from the $x$ and $y$ components of the Wald-Zoupas CM angular momentum (see Appendix~\ref{app:STF_spherical_harmonics_conversion}).

\subsubsection{Post-Newtonian results}

For nonprecessing binaries, the mass and current multipole moments $U_{lm}$ and $V_{lm}$ are expressed conveniently in terms of several different mass parameters and mass ratios. 
Here we denote the individual masses by $m_1$ and $m_2$ with $m_1 > m_2$. 
We then denote the total mass by $M=m_1+m_2$, the relative mass difference by $m_{12} = (m_1-m_2)/M$, the mass ratio by $q = m_1/m_2 \geq 1$, and the symmetric mass ratio $\nu=m_1m_2/M^2$.
We also use the notation $\Omega$ for the orbital frequency, $\psi$ for the orbital phase, and $x=(M \Omega)^{2/3}$ for the PN parameter, as in~\cite{Blanchet:2013haa}. 
It is shown in~\cite{Blanchet:2013haa} that all the waveform modes $h_{lm}$ can be written in the form
\begin{equation}
    h_{lm} = -\frac{8 M\nu x}{r} \sqrt{\frac \pi 5} \mathcal H_{lm} e^{-im\psi} \, ,
\end{equation}
where the terms $\mathcal H_{lm}$ are given in Eqs.~(328)--(329) of~\cite{Blanchet:2013haa} and can be written as polynomials in the square root of the PN parameter (i.e., $\sqrt x$).
We do not use the full expressions for $\mathcal H_{lm}$ in Eqs.~(328)--(329) of~\cite{Blanchet:2013haa}; rather we only go up to 2.5PN order (i.e., $x^{5/2}$) in these equations.
After substituting these expressions into Eq.~\eqref{eq:delta_J10_radiative_multipole_moments}, we find that the result for $\delta J_{1,0}^{\alpha=\beta=1}$ is given by
\begin{align}
\label{eq:deltaJ10_PN}
    \delta J_{10}^{(\alpha=\beta=1)} = {} & \frac{8}{5}\sqrt{\frac{3\pi}{2}} M^2 \nu^2 \left(-\frac{10}{21}-\frac{m_{12}^2}{210}+\frac{9329}{4410}\nu \right)x^{9/2} \nonumber \\
    & + O(x^5) \, .
\end{align}
The angular momentum in the Newtonian limit goes as $x^{-1/2}$, so the correction term in Eq.~\eqref{eq:deltaJ10_PN} appears at 5PN order with respect to the leading-order effect.
During the inspiral when the PN parameter $x$ is small, $\delta J_{10}^{\alpha=\beta=1}$ is not expected to be very large. 
Given the fact that the product $\bar U_{22} V_{32}$ scales with the PN parameter as $x^3$, it might initially seem unusual that the net effect $\delta J_{10}^{\alpha=\beta=1}$ goes like $x^{9/2}$.
Because there is a real part in Eq.~\eqref{eq:delta_J10_radiative_multipole_moments}, there are a number of cancellations that occur between different modes.
These cancellations in the $U_{lm}$ and $V_{lm}$ moments occur in the conservative part of the dynamics, but not the dissipative part from GW radiation reaction. 
These dissipative dynamics appear as a relative 1.5PN correction to $V_{32}$, which explains why the leading order part of $\delta J_{10}^{\alpha=\beta=1}$ goes like $x^{9/2}$.
Analogous arguments can be made for the other terms in Eq.~\eqref{eq:delta_J10_radiative_multipole_moments}.

There is another feature of Eq.~\eqref{eq:deltaJ10_PN} worth describing that relates to the dependence of $\delta J_{10}^{\alpha=\beta=1}$ on the mass ratio $q$ (and which is a feature that also appears in the NR simulations, which we discuss later). 
Specifically, the sign of $\delta J_{10}^{\alpha=\beta=1}$ changes, and there is a specific mass ratio at which the leading PN expression vanishes. 
The value of the mass ratio can be computed from Eq.~\eqref{eq:deltaJ10_PN} to be $q\approx 1.9$.
The physical reason for this value was less clear to us, though it arises from the change in amplitudes of the multipole moments $U_{lm}$ and $V_{lm}$ as a function of mass ratio $q$.

The leading-order contribution to $\delta k_{1\pm1}^{(\alpha=\beta=1)}$ turns out to require fewer terms to compute, as indicated in Eq.~\eqref{eq:delta_k_radiative_multipole_moments}, and it only requires the leading-order parts of the moments $U_{22}$ and $V_{21}$.
It is reasonably straightforward to show that $\delta k_{1\pm1}^{(\alpha=\beta=1)}$ is given by
\begin{align}
\label{eq:delta_k_PN}
    \delta k_{1,\pm1}^{(\alpha=\beta=1)} = - i \frac{22}{35}\sqrt{\frac{\pi}{3}} M^2 \nu^2 m_{12} x^{5/2} e^{\mp i\psi} + O(x^3) \, .
\end{align}
The difference term from the Wald-Zoupas definition of the CM angular momentum scales as $x^{5/2}$, which is two PN orders lower than the correction term to the intrinsic angular momentum.
However, this effect also goes as $e^{\mp i\psi}$, so the average over an orbital period vanishes.
As was discussed in~\cite{Nichols:2018qac}, while the change in the Wald-Zoupas definition of the CM angular momentum scales with the PN parameter as $x^0 = O(1)$, there is a choice of reference time $u_0$ that can set the change in the CM angular momentum to zero through 2PN order (i.e., through $x^2$).
At 2.5PN order ($x^{5/2}$), there is no longer just a choice of reference time that allows the effect to be set to zero, which also preserves the fact that the binary was initially chosen to be in the CM frame and rest-frame of the source with the supertranslations chosen such that $C_{AB} = 0$ initially.
Thus, the terms $\delta k_{1\pm1}^{(\alpha=\beta=1)}$ in Eq.~\eqref{eq:delta_k_PN} are of the same PN order as the nontrivial (in the sense discussed here) Wald-Zoupas CM angular momentum.
The impact of the different definitions of angular momentum is thus largest for the CM angular momentum (although the impact of the CM angular momentum on the evolution of compact binaries has not been discussed as extensively as that of the other charges associated with the Poincar\'e group).

Finally, we also point out that from Eq.~\eqref{eq:delta_k_PN} it can be shown that the maximum effect happens approximately at $q=2.6$. 
This is comparable to the value of the mass ratio that results in the maximum kick velocity for nonspinning binaries ($q=2.8\pm 0.23$)~\cite{Gonzalez:2006md}.

\subsubsection{Results from NR surrogate models}

While the PN approximation gives useful intuition about the effect of the remaining free parameter $\alpha$ on the intrinsic and CM angular momentum during the inspiral phase of a compact binary, it is not expected to be accurate during the merger and ringdown phases.
Instead, it is preferable to use the results of NR simulations during these late stages of a BBH merger.
In particular, we will use the hybrid NR surrogate model NRHyb3dq8~\cite{Varma:2018mmi} to generate the waveform modes that enter into Eqs.~\eqref{eq:delta_J10_radiative_multipole_moments} and~\eqref{eq:delta_k_radiative_multipole_moments}. 
The surrogate produces the waveform modes $r h_{lm}/M$, which we convert to the $U_{lm}$ and $V_{lm}$ moments using Eq.~\eqref{eq:htoUVlm}. 
Because the surrogate does not model the modes $h_{40}$, $h_{41}$ and $h_{53}$, we cannot include the surrogate model's contribution to these modes in Eq.~\eqref{eq:delta_J10_radiative_multipole_moments}.
Also, because the surrogate does not have the memory or spin memory contributions to the modes $h_{20}$, $h_{30}$, and $h_{40}$, we add these contributions to those of the surrogate model.
The procedure we use to compute these memory modes is reviewed in Sec.~\ref{subsec:MemoryModes}.

For presenting our results from the surrogate waveforms, we opt to show the Cartesian components of the intrinsic or CM angular momentum instead of the multipole moments that were described in the previous parts. 
The conversion between these two descriptions is reasonably straightforward and is described in further detail in Appendix~\ref{app:STF_spherical_harmonics_conversion}.
We thus quote the results here.
First, the $z$ component for the intrinsic angular momentum $\delta J_z^{(\alpha=\beta=1)}$ can be related to $\delta J_{10}^{(\alpha=\beta=1)}$ by
\begin{equation} \label{eq:deltaJz}
    \delta J_z^{(\alpha=\beta=1)} = -2 \sqrt{\frac{2\pi}{3}} \delta J_{10}^{(\alpha=\beta=1)} \, .
\end{equation}
Similarly, $\delta k_x^{(\alpha=\beta=1)}$ and $\delta k_y^{(\alpha=\beta=1)}$ can be related to $\delta k_{1\pm1}^{(\alpha=\beta=1)}$ by
\begin{subequations}
\label{eq:deltakxy}
\begin{align}
    \delta k_x^{(\alpha=\beta=1)} = {} & 
    -4\sqrt{\frac{\pi}{3}}\Re\left[\delta k_{11}^{(\alpha=\beta=1)}\right] \, ,\\
    \delta k_y^{(\alpha=\beta=1)} = {} & 
    4\sqrt{\frac{\pi}{3}}\Im\left[\delta k_{11}^{(\alpha=\beta=1)}\right]
\end{align}
\end{subequations}
(see also~\cite{Nichols:2018qac}).
Because $\delta k_z^{(\alpha=\beta=1)}$ is proportional to $\delta k_{10}^{(\alpha=\beta=1)} = 0$ for nonspinning BBHs, then the magnitude of the difference of the CM angular momentum is given by
\begin{equation}
    |\delta \mathbf{k}^{(\alpha=\beta=1)}| = \sqrt{\left(\delta k_x^{(\alpha=\beta=1)}\right)^2 + \left(\delta k_y^{(\alpha=\beta=1)}\right)^2} \, .
\end{equation}

\begin{figure}[t!]
    \centering
    \includegraphics[width=\columnwidth]{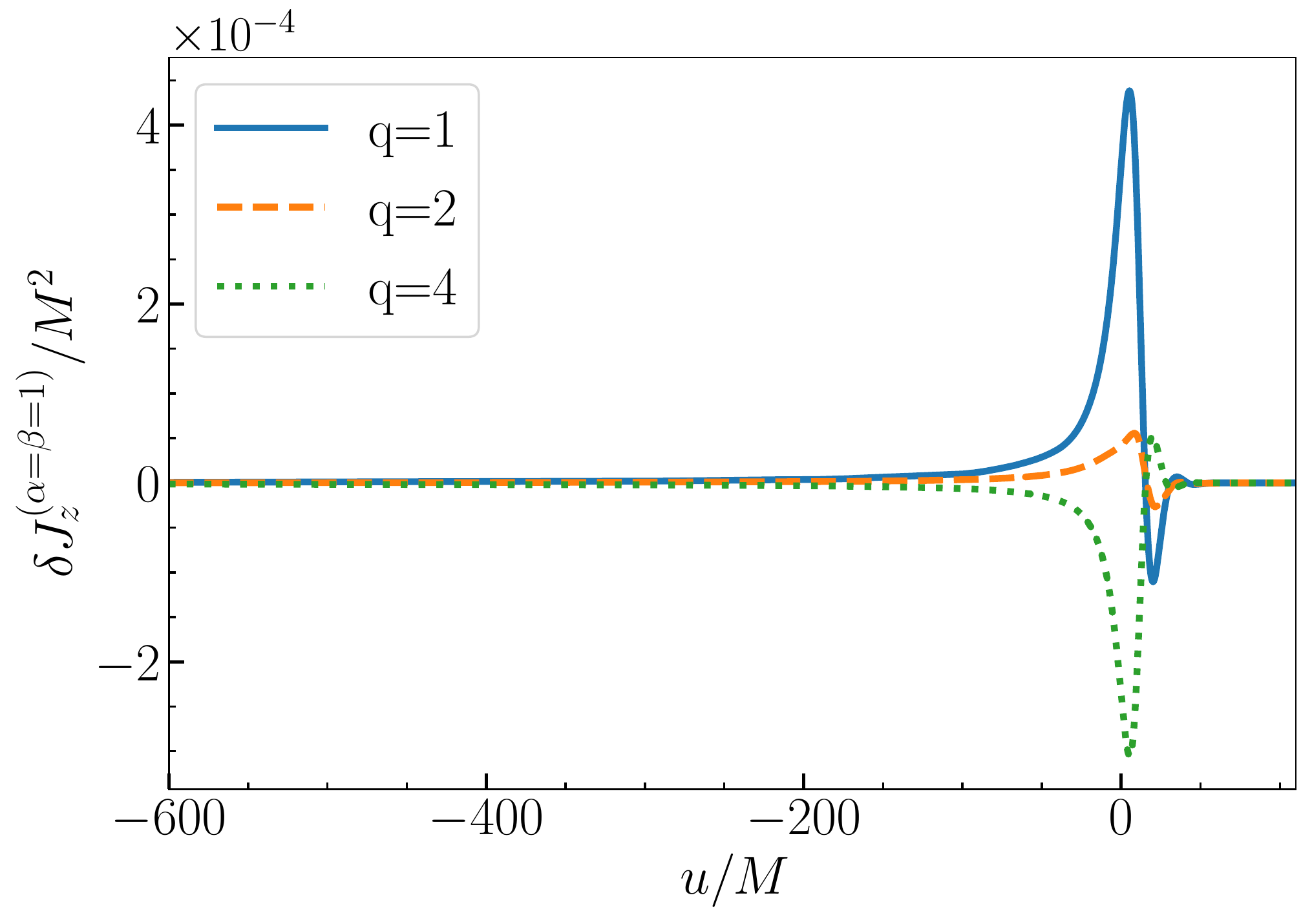}
    \includegraphics[width=\columnwidth]{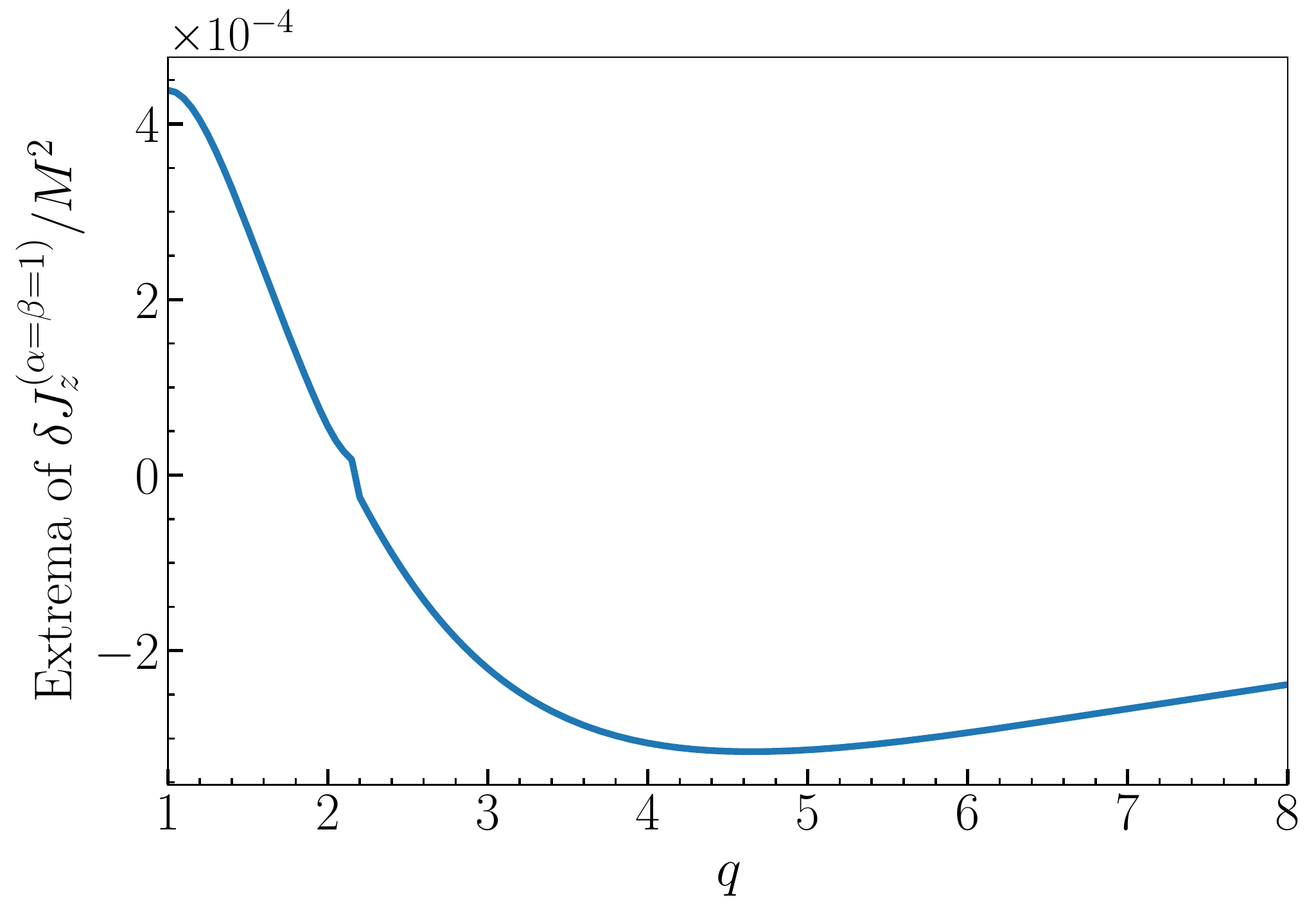}
    \caption{\textit{Top}: The $z$ component of the difference of the intrinsic angular momentum from the Wald-Zoupas values (denoted by $\delta J_z^{(\alpha=\beta=1)}$) as a function of retarded time for nonspinning BBH mergers of three mass ratios, $q=1$, 2, and 4.
    Note that the extreme value switches from a maximum to a minimum as a function of mass ratio.
    As discussed further in the text, $\delta J_z^{(\alpha=\beta=1)}$ was computed using a NR surrogate model (where the peak of the magnitude of the waveform is at retarded time equal to zero) using Eqs.~\eqref{eq:delta_J10_radiative_multipole_moments} and~\eqref{eq:deltaJz}.
    \textit{Bottom}: The extreme value of the $z$ component of $\delta J_z^{(\alpha=\beta=1)}$ as a function the mass ratio.
    Consistent with the PN predictions, there is a change in the sign of the quantity $\delta J_z^{(\alpha=\beta=1)}$ that occurs near the mass ratio $q=2$.}
    \label{fig:Jz-q}
\end{figure}

We first show the difference of the intrinsic angular momentum from the Wald-Zoupas value, $\delta J_z^{(\alpha=\beta=1)}$, for BBHs with different mass ratios. 
The top panel of Fig.~\ref{fig:Jz-q} displays $\delta J_z^{(\alpha=\beta=1)}$ as a function of retarded time for three different mass ratios, $q=1$, 2, and 4 as solid blue, orange dashed, and green dotted curves, respectively.
The extreme values of the time series for $\delta J_z^{(\alpha=\beta=1)}$ approach the largest positive, the closest to zero, and the most negative value for these three mass ratios, respectively.
The dependence of the extreme value of $\delta J_z^{(\alpha=\beta=1)}$ as a function of mass ratio is illustrated in more detail in the bottom panel of Fig.~\ref{fig:Jz-q}. 
As was noted in the discussion of $\delta J_z^{(\alpha=\beta=1)}$ in the PN approximation, the extreme value of this quantity changes sign as a function of mass ratio. 
The value at which it undergoes this sign change for the surrogate model is $q \approx 2.2$, which is close to the value predicted by the leading PN result of $q\approx 1.9$.
There is a sharp feature in the curve near the mass ratio where $\delta J_z^{(\alpha=\beta=1)}$ goes to zero, because (what is for most mass ratios) the primary peak (which changes smoothly with mass ratio) becomes smaller than (what is for most mass ratios) the secondary peak (which also varies smoothly with mass ratio, but at a different rate from the primary peak).
When the roles of primary and secondary peak reverse for a small range of mass ratios, the slope changes abruptly, and this leads to this slight sharp feature.

We also mention a few implications of the results presented in Fig.~\ref{fig:Jz-q}. 
During the inspiral, the Newtonian value of the orbital angular momentum is given by $M^2 \nu x^{-1/2}$.
For an equal mass binary separated by a distance of $100 M$, the angular momentum will initially be of order $\sim 2.5 M^2$.
The final black hole is a Kerr black hole with spin of order $\sim 0.67 M_\mathrm{f}^2$, where $M_\mathrm{f}$ is the final mass of the black hole (which is typically at least ninety percent of the total mass $M$).
Thus, the fact that $\delta J_z^{(\alpha=\beta=1)}$ is of order a few times $10^{-4} M^2$ at its largest implies that the discrepancies in the definitions of angular momentum will be small for definitions where $\alpha$ is of order unity.
However, the final spin parameter of the black hole formed from a BBH merger is often quoted to an accuracy which is smaller than the values of $\delta J_z^{(\alpha=\beta=1)}$ described here (see, e.g.,~\cite{Boyle:2019kee}).
Thus, for completeness, NR simulations should specify which definition of angular momentum is being used.

\begin{figure}[t!]
    \centering
    \includegraphics[width=\columnwidth]{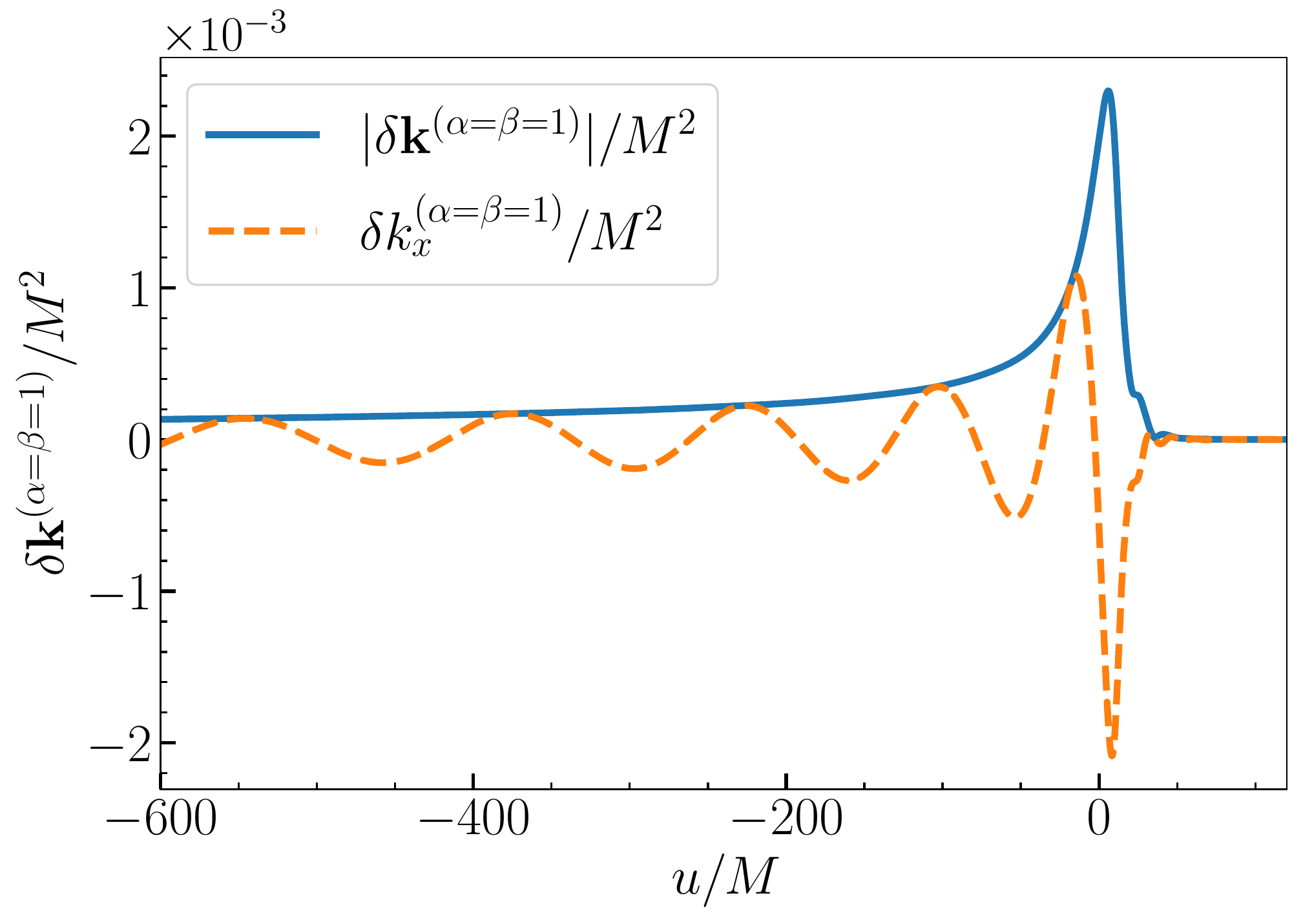}
    \includegraphics[width=\columnwidth]{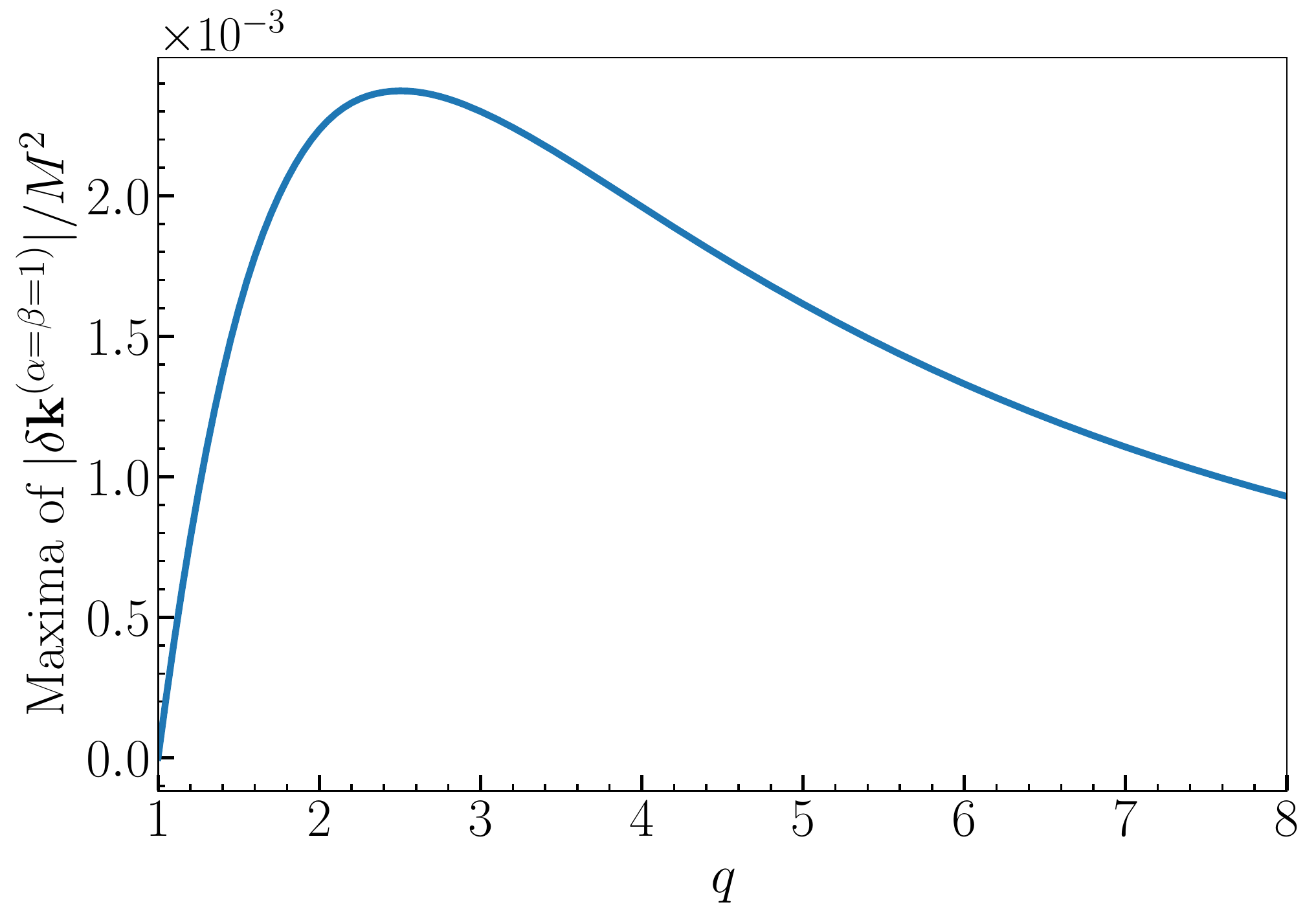}
    \caption{\textit{Top}: The magnitude and the $x$ component of the difference of the CM angular momentum from the Wald-Zoupas definition, $|\delta \mathbf{k}^{(\alpha=\beta=1)}|$ and $\delta k_x^{(\alpha=\beta=1)}$, respectively, as functions of retarded time.
    The system shown is a BBH merger with mass ratio $q=3$, and the waveform modes used in Eqs.~\eqref{eq:delta_k_PN} and~\eqref{eq:deltakxy} were generated from a NR surrogate, where the peak magnitude of the waveform occurs at a time equal to zero.
    The vector $\delta \mathbf{k}^{(\alpha=\beta=1)}$ is in phase with the orbital motion of the binary during inspiral, and it grows in magnitude until the merger, after which it settles to zero.
    \textit{Bottom}: The maximum of the magnitude of the difference of the CM angular momentum from the Wald-Zoupas value as a function of the mass ratio of a BBH system. 
    Note that the maximum value as a function of $q$ occurs at roughly the same mass ratio that produces the maximum kick velocity of the final black hole (see the text for further discussion).}
    \label{fig:kvsq}
\end{figure}

We now turn to the difference of the CM angular momentum from the Wald-Zoupas value.
We use the same surrogate model to compute $\delta k_x^{(\alpha=\beta=1)}$ and $|\delta \mathbf{k}^{(\alpha=\beta=1)}|$ as functions of retarded time. 
We plot these quantities in the top panel of Fig.~\ref{fig:kvsq} for $q=3$. 
The bottom panel of Fig.~\ref{fig:kvsq} shows the peak value of the time series $|\delta \mathbf{k}^{(\alpha=\beta=1)}|$ as a function of the binary's mass ratio, $q$. 
For an equal mass black-hole binary, $q=1$, the change in the CM angular momentum vanishes.
This occurs because there is no linear momentum radiated from such a system, so the initial and final rest frames are the same (and we have chosen the initial rest frame to be the CM frame). 
The peak value of $|\delta \mathbf{k}^{(\alpha=\beta=1)}|$ is reached at a mass ratio of roughly $q \approx 2.5$.
This is similar to the PN prediction of $q\approx 2.6$ computed earlier.
It is also near the peak value of the gravitational recoil computed in~\cite{Gonzalez:2006md} of $q\approx 2.8$.
The decrease in the magnitude of $|\delta \mathbf{k}^{(\alpha=\beta=1)}|$ at mass ratios greater than $q \sim 2.5$ is likely related to the fact that the gravitational recoil also decreases at these larger mass ratios.

As far as we are aware, there has not been a systematic study of the size Wald-Zoupas CM angular momentum from numerical relativity simulations.
In the PN approximation, the calculations in~\cite{Nichols:2018qac}, which were reviewed in this subsection, suggest that the magnitude of the Wald-Zoupas CM angular momentum, $|\mathbf{k}^{(\alpha=\beta=1)}|$, goes as $M^2 x^{5/2}$.
Thus, the magnitude of the CM angular momentum could be as large as order $M^2$ near the merger (thereby making the difference $|\delta \mathbf{k}^{(\alpha=\beta=1)}|$ a small effect).
Further investigation is needed to have a more definitive statement about the possible importance of the term $|\delta \mathbf{k}^{(\alpha=\beta=1)}|$.

\subsection{Super angular momentum}

We now turn to understanding effect of the free parameter $\alpha$ ($=\beta$) on the difference of the super angular momentum from the charge of~\cite{Compere:2018ylh} for nonspinning BBH mergers.
Unlike the angular momentum, the super angular momentum can have a nontrivial net change between the early- and late-time nonradiative regions of a spacetime for these systems.
We thus focus on the net change in the charges $\Delta Q_Y^{\alpha=\beta}$: namely, the difference of Eq.~\eqref{eq:QYalpha_equals_beta} between two nonradiative regions at early and late times.
Thus, we will similarly be interested in the change in the difference term from the $\alpha=\beta=1$ value of the charges; i.e., the quantity $\Delta \delta Q_Y^{\alpha=\beta=1}$, where $\delta Q_Y^{\alpha=\beta=1}$ is defined in Eq.~\eqref{eq:deltaQYalpha_equals_beta}.

We now calculate the change in the largest (in magnitude) nonvanishing part of the super angular momentum, which appears in the $l=2$, $m=0$ moments of the super-CM part (in both the PN approximation and from NR simulations).
First, we write the expression for this change in the charges as
\begin{equation}
\Delta Q_{(e),20}^{(\alpha=\beta)} = \Delta Q_{(e),20}^{(\alpha=\beta=1)}+ (\alpha-1) \Delta \delta Q_{(e),20}^{(\alpha=\beta=1)} \, .
\end{equation}
The change in the term $\delta Q_{(e),20}^{(\alpha=\beta=1)}$ can be obtained by taking the difference of Eq.~\eqref{eq:superCMalphaEQbeta} evaluated at early and late times. 
For nonspinning binaries, all the $V_{lm}$ moments vanish in nonradiative regions; the change in the moments $U_{lm}$ can be nonvanishing in nonradiative regions when there is a nontrivial GW memory effect.
The largest moments are $U_{20}$ and $U_{40}$, as described in Sec.~\ref{subsec:MemoryModes}; however, because the mode $U_{40}$ is a factor of $60\sqrt{3}$ times smaller than the $U_{20}$ mode, we focus here on the contribution from just $U_{20}$.
We find that the leading change in the difference term is given by
\begin{equation} \label{eq:DeltadeltaQ20}
    \Delta \delta Q_{(e),20}^{(\alpha=\beta=1)} = \frac{3}{448 \pi}\sqrt{\frac{15}{2 \pi}} \Delta (U_{20})^2 \, .
\end{equation}
Finally, we will compute $\Delta Q_{(e),20}^{(\alpha=\beta=1)}$.
The term quadratic in $C_{AB}$ in Eq.~\eqref{eq:QYalphabetaCharge} gives rise to a term quadratic in $\Delta U_{20}$ which is identical to the expression for $\Delta \delta Q_{(e),20}^{(\alpha=\beta=1)}$ in Eq.~\eqref{eq:DeltadeltaQ20}.
The term linear in the shear does not contribute (because it involves only $V_{lm}$ modes) and the term $- u D_A m$ does not have a contribution from nonspinning BBH mergers to this part of the charge.
However, the term involving $N_A$ in Eq.~\eqref{eq:QYalphabetaCharge} does contribute to $\Delta Q_{(e),20}^{(\alpha=\beta=1)}$.
The form of $N_A$ is known in stationary regions that are supertranslated from the canonical frame in which $C_{AB}=0$.
It was shown in~\cite{Flanagan:2015pxa} that $N_A = - 3 m D_A \Phi/2$, where $\Phi$ is the ``potential'' for the electric part of the shear [as in Eq.~\eqref{eq:canonical_frame_shear}], and the Bondi mass aspect $m$ is a constant in this frame.
Using the fact that $\Delta U_{20} = \sqrt{12} \Delta \Phi_{20}$, we then find that the leading $\alpha=\beta=1$ super CM is given by
\begin{equation} \label{eq:super_CM_charge_final}
    \Delta Q_{(e),20}^{(\alpha=\beta)} = \frac{-3}{16\pi}\frac{M}{\sqrt{2}} \Delta U_{20}+ \frac{3}{448 \pi}\sqrt{\frac{15}{2 \pi}} \Delta (U_{20})^2 \, .
\end{equation}
The lowest multipole moment (consistent with the symmetries of nonprecessing BBHs) in which the change in the superspin part could appear is the $l=3$, $m=0$ mode.
When we evaluate the contribution of the $U_{20}$ modes in Eq.~\eqref{eq:deltaQalphab} for $l=3,m=0$, we find it and the difference from the Hamiltonian charge of~\cite{Compere:2018ylh} both vanish:
\begin{equation}
    \Delta Q_{(b),30}^{(\alpha=\beta=1)} = \Delta \delta Q_{(b),30}^{(\alpha=\beta=1)} = 0 \, .
\end{equation}
Note, however, that the instantaneous value of the charges (not the change in a nonradiative-to-nonradiative transition) can be nonvanishing, though we do not compute that quantity here.
We next turn to the computation of the super CM using the PN approximation and the NR surrogate model discussed in the previous subsection.

\paragraph*{PN approximation}
We calculate the $U_{20}$ waveform modes associated with the GW memory effect as was described in Sec.~\ref{subsec:MemoryModes}.
Because the PN approximation covers only the inspiral, we truncate the calculation of $\Delta U_{20}^{(\alpha=\beta=1)}$ at a finite retarded time $u$, at which the binary is at a PN parameter $x$.
We thus denote the change in the PN parameter by $\Delta x$.
This gives an expression for the $U_{20}$ moment that is equivalent to the one given in~\cite{Blanchet:2013haa}.
We thus find that the change in the super-CM angular momentum in Eq.~\eqref{eq:super_CM_charge_final} and the change in the difference in Eq.~\eqref{eq:DeltadeltaQ20} are given by
\begin{subequations}
\begin{align}
    & \Delta Q_{(e),20}^{(\alpha=\beta)} = {}  
    -\frac{1}{28}\sqrt{\frac{15}{2\pi}}M^2\nu \Delta x +\frac{5}{1372}\sqrt{\frac{15}{2\pi}} M^2 \nu^2 \Delta (x^2) \, ,\\
    & \Delta \delta Q_{(e),20}^{(\alpha=\beta=1)} = {}  \frac{5}{1372}\sqrt{\frac{15}{2\pi}} M^2 \nu^2 \Delta (x^2) \, .
\end{align}
\end{subequations}
Thus, the different definitions of the super-CM angular momentum causes a relative 1PN-order correction to the leading-order super-CM angular momentum.

\paragraph*{Numerical-relativity results}
The GW memory effect is largest not during the inspiral, but after the merger and ringdown of a BBH collision.
To better understand the size of the change in the super-CM angular momentum of a BBH merger, we compute the full memory effect in the $U_{20}$ mode as in Eq.~\eqref{eq:U20memory}, and we substitute the result into Eqs.~\eqref{eq:DeltadeltaQ20} and~\eqref{eq:super_CM_charge_final}.
We again consider nonspinning BBH mergers of different mass ratios, and we use the same hybrid surrogate model NRHybSur3dq8~\cite{Varma:2018mmi} to compute $\Delta U_{20}$.
We take the mass $M$ that enters into Eq.~\eqref{eq:super_CM_charge_final} to be the final mass, which we compute using the NR fits computed in~\cite{Varma:2018aht}.

\begin{figure}[t!]
    \centering
    \includegraphics[width=\columnwidth]{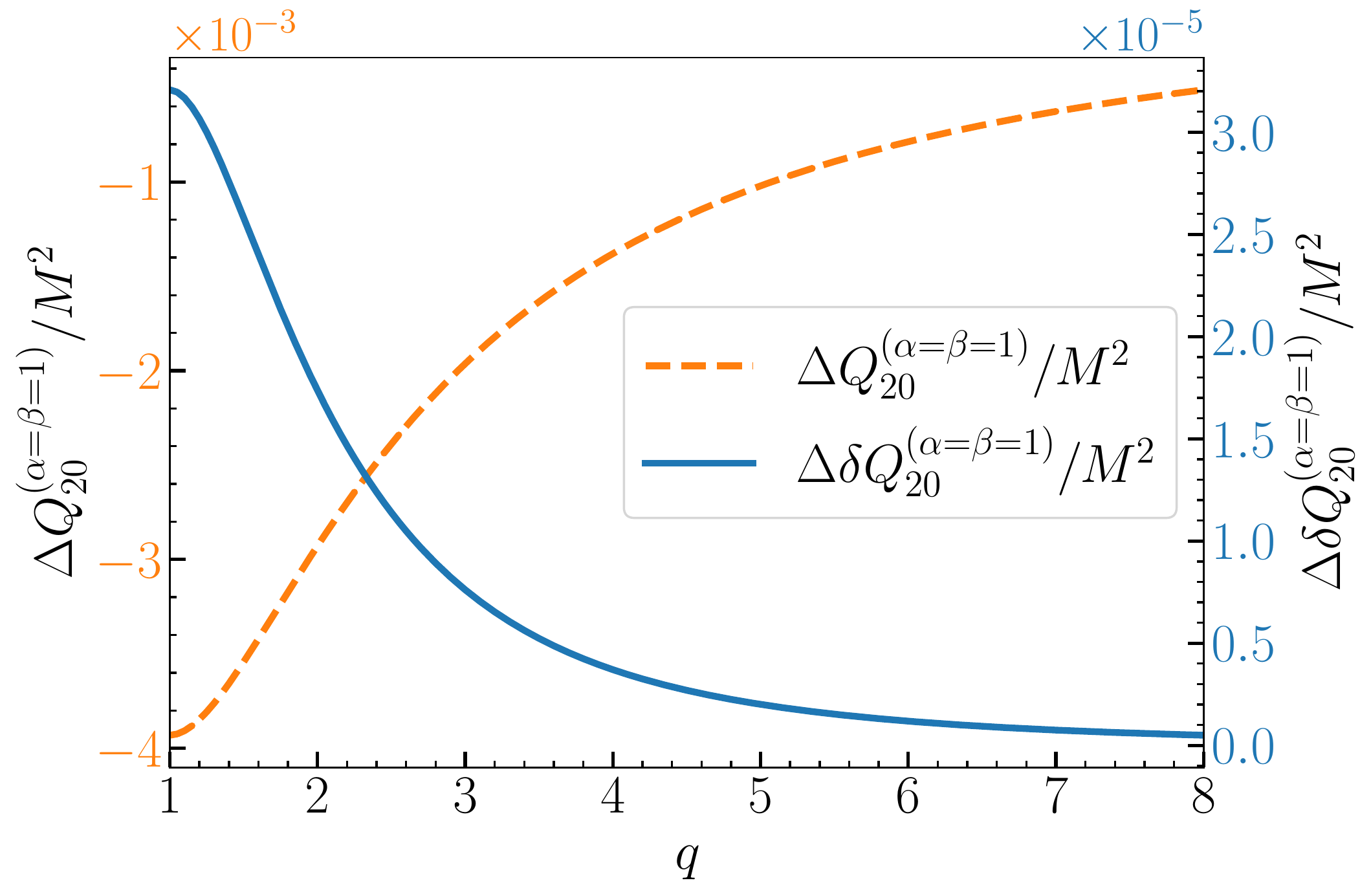}
    \caption{The change in the Hamiltonian super-CM angular momentum of~\cite{Compere:2018ylh}, $\Delta Q_{2,0}^{(\alpha=\beta=1)}$ (scale on the left), and the change in the difference of the super-CM angular momentum from the Hamiltonian super-CM angular momentum of~\cite{Compere:2018ylh}, $\Delta\delta Q_{2,0}^{(\alpha=\beta=1)}$ (scale on the right), both as a function of the mass ratio of the binary $q$.
    The difference term is about two orders of magnitude smaller that the change in the super CM.}
    \label{fig:dQ20vsq}
\end{figure}

In Fig.~\ref{fig:dQ20vsq}, we show the net change in difference in the super-CM angular momentum from the Hamiltonian super-CM angular momentum of~\cite{Compere:2018ylh}, as a function of the mass ratio of nonspinning BBH mergers of different mass ratios between $1\leq q\leq 8$. 
The maximum difference occurs for equal-mass BBHs and decreases with higher mass ratios, which is consistent with the amplitude of the memory effect computed from the dominant quadrupole modes, as in Eq.~\eqref{eq:U20memory}.
This figure illustrates that the change in the difference terms of the leading super-CM angular momentum are about one hundredth of the change in the super-CM of~\cite{Compere:2018ylh}, which is itself a small effect in units of $M^2$.
Nevertheless, the waveform modes used to compute the result are sufficiently accurate that this difference can be resolved.

\section{\label{sec:Conclusions} Conclusions}

In this paper, we investigated the freedoms in defining angular momentum and super angular momentum in asymptotically flat spacetimes and the implications of these freedoms on the values of the (super) angular momentum of nonspinning binary-black-hole mergers.
The fact that such freedoms exist was recently discussed in~\cite{Compere:2019gft}, which demonstrated that there can be a two (real) parameter family of angular momenta, which encompass a few commonly used definitions of angular momentum in asymptotically flat spacetimes. 
All members of this two-parameter family satisfy flux balance laws and are constructed from quantities that are covariant with respect to 2-sphere cross sections of null infinity.
We found, however, that for the angular momentum to vanish in flat spacetime, the two parameters must be equal; this leads to a natural requirement that the family of angular momenta should depend upon only a single real parameter.
If we do not require that the angular momentum agree with the Hamiltonian definition of Wald and Zoupas, then there remained a one-parameter family of angular momentum.

We further investigated the effect of this one free parameter on the values of the angular momentum. 
To do so, we first derived a multipolar expansion (in terms of the radiative multipole moments of the GW strain) of the difference of the angular momentum from the Wald-Zoupas definition. 
The difference is constructed from the products of mass moments with current moments, unlike the flux of the Wald-Zoupas definition of angular momentum, which is written in terms of products of mass moments with themselves and current moments with themselves. 
This fact has an important implication for spacetimes that transition between nonradiative regions at early times and at late times, the context in which the GW memory effect is usually computed.
For several types of systems of astrophysical interest, such as compact-object mergers, the GW memory effect appears in just the mass-type moments.
Thus, the difference terms that arise from products of mass and current moments will vanish in these nonradiative-to-nonradiative transitions, and the net change in the angular momentum will be independent of this remaining free parameter.
There will, however, be a difference in the instantaneous value of the angular momentum while the system is radiating gravitational waves.

We also proposed considering a two-parameter family of super angular momentum in analogy with the two-parameter family of angular momentum given in~\cite{Compere:2019gft}. 
Choosing the two parameters to be equal does not generically make the super angular momentum vanish in flat spacetime (and it has also been argued that the super angular momentum should not necessarily vanish in this context).
There is a choice of the two parameters that does manifestly make the super angular momentum vanish in flat spacetime, but it does not correspond to the analog of the Wald-Zoupas charge. 
We, therefore, derived a multipolar expansion of the difference in the super angular momentum from the Hamiltonian definition of~\cite{Compere:2018ylh} that involved two real parameters. 
We also specialized the result to have one free parameter, so that the charge reduces to the angular momentum when the symmetry vector field reduces from an infinitesimal super Lorentz transformation to a standard infinitesimal Lorentz transformation.
 
Next, we investigated the magnitude of the difference of the (super) angular momentum from the Wald-Zoupas charges for nonspinning, quasicircular binary-black-hole mergers. 
For the standard angular momentum the difference occurs only while the system is radiating GWs.
In the post-Newtonian approximation, we found the difference in the intrinsic angular momentum enters at a relative 5PN-order to the Newtonian angular momentum, while the difference in the CM angular momentum, it appears at the same PN order as the effect that cannot be set to zero through a particular choice of reference time (at 2.5PN order beyond the leading Newtonian expression). 
Given the high PN orders, the effects will generally be small, although they could become large near the binary's merger, when the PN approximation becomes inaccurate.
During the inspiral, however, the difference in the CM angular momentum from the Wald-Zoupas value will be larger than that of the intrinsic angular momentum, because of its lower PN order. 
For the super angular momentum, the difference terms need not vanish after the radiation passes; thus, we focused on the net change of the charges between early times and late times.
We found that the leading difference in the superspin vanishes for BBH mergers, while differences in the super-CM angular momentum cause a relative 1PN difference from the Hamiltonian super-CM angular momentum of~\cite{Compere:2018ylh}.

Finally, we estimated the difference terms for the (super) angular momentum using inspiral-merger-ringdown surrogate waveforms of nonspinning BBH mergers that were fit to numerical-relativity simulation data. 
The intrinsic angular momentum terms are largest at equal mass, change sign at a mass ratio near two, and then take on the most negative value near a mass ratio of four before approaching closer to zero.
The amplitude of the effect is small compared to the Newtonian value of the angular momentum.
The maximum difference in the CM angular momentum was found to happen approximately at the mass ratio that produces the maximum kick velocity of the final black hole. 
The difference in the change of the super-CM angular momentum from the corresponding Hamiltonian expression of~\cite{Compere:2018ylh} in a nonradiative-to-nonradiative transition was only to a few percent correction.
Although these differences in the (super) angular momentum are small compared to the values of the (super) angular momentum itself, they are able to be resolved for these systems.
Thus, which definition is being used should be specified when describing the (super) angular momentum of nonspinning binary-black-hole mergers.

\begin{acknowledgments}
A.E.\ and D.A.N.\ acknowledge support from the NSF grant PHY-2011784. 
We thank Alex Grant for helpful discussions about the Wald-Zoupas definition of angular momentum in the covariant conformal approach to null infinity and for comments on the manuscript.
We thank Geoffrey Comp\`ere and Ali Seraj for correspondence related to their work~\cite{Compere:2019gft}; we also thank Geoffrey Comp\`ere, Adrien Fiorruci, and Romain Ruzziconi for correspondence related to their work~\cite{Compere:2020lrt}.
\end{acknowledgments}

\appendix
\section{\label{app:STF_spherical_harmonics_conversion} Conversion between STF tensors and spherical harmonics}

In this section, we compare our expressions for the difference in the intrinsic and center-of-mass angular momentum from the Wald-Zoupas values in Eqs.~\eqref{eq:intrinsic_angular_momentum} and~\eqref{eq:CM_angular_momentum} to a related result obtained by Comp\`ere \textit{et al.} in~\cite{Compere:2019gft}.
We start with the intrinsic angular momentum terms, and we make this comparison by converting the $u$ integral of the expression in Eq.~(4.16) of~\cite{Compere:2019gft} for the intrinsic angular momentum in terms of STF $l$-index tensors $U_L \equiv U_{\langle i_1\ldots i_l \rangle}$ and $V_L \equiv V_{\langle i_1\ldots i_l \rangle}$ to the multipole moments $U_{lm}$ and $V_{lm}$ used in this paper (the angle brackets around indices mean that the symmetric, trace-free part of the tensor should be taken).
We focus on the second term in Eq.~(4.16) of~\cite{Compere:2019gft} which represents the difference from the Wald-Zoupas value of the angular momentum. 
We denote this correction term by $\delta J_i^{(\alpha=\beta=1)}$, 
where the index $i$ means the angular momentum was computed with respect to a vector on the 2-sphere $Y^A_i = \epsilon^{AB} D_B n_i$.
The quantity $n_i$ is a unit vector in quasi-Cartesian coordinates that is constructed from spherical polar coordinates $(\theta,\phi)$ as follows
\begin{equation} \label{eq:niCartesian}
    n_i=(\sin\theta \cos\phi, \sin\theta \sin\phi, \cos\theta) \, .
\end{equation}
The expression for $\delta J_i^{(\alpha=\beta=1)}$ from~\cite{Compere:2019gft} is given by
\begin{equation}
\label{eq:ang_mom_STF}
    \delta J_i^{(\alpha=\beta=1)} = -\sum_{l\geq 2} (l+1)^2 \mu_{l+1} \left(b_l U_{iL} V_L - b_{l+1} U_L V_{iL}\right) \, .
\end{equation}
The coefficients $b_l$ (not to be confused with $b^{(\pm)}_{lm}$ defined in the main text) and $\mu_l$ were defined in~\cite{Compere:2019gft} to be
\begin{subequations}
\begin{align}
    b_l = {} & \frac{2l}{l+1} \, ,\\
    \mu_l={} & \frac{(l+1)(l+2)}{(l-1)ll!(2l+1)!!} \, .
\end{align}
\end{subequations}
To rewrite Eq.~\eqref{eq:ang_mom_STF} in terms of $U_{lm}$ and $V_{lm}$ modes, we relate the spherical harmonics $Y^{lm}$ to the symmetric trace-free tensors of rank-$l$ (STF-$l$ tensors) $N_L=n_{\langle i_1}\ldots n_{i_l\rangle }$ using the result in~\cite{RevModPhys.52.299}
\begin{equation}
\label{eq:STF_spherical}
    Y^{lm} = \mathcal{Y}^{lm}_L N_L \, .
\end{equation}
The tensors $\mathcal{Y}_L^{lm}$ with $-l\leq m \leq$ are a basis for the vector space of $l$-index STF tensors and are defined in~\cite{RevModPhys.52.299} (we do not need their explicit form here).
They transform under complex conjugation in the same way as the scalar spherical harmonics: 
\begin{equation}
\label{eq:conj_STF}
    \bar{\mathcal{Y}}^{lm}_L = (-1)^m \mathcal{Y}^{l,-m}_L \, .
\end{equation}

The STF mass and current moments $U_L$ and $V_L$ are related to  $U_{lm}$, $V_{lm}$, and $\mathcal{Y}_L^{lm}$ by
\begin{subequations}
\begin{align}
    U_L&=\frac{l!}{4}\sqrt{\frac{2l(l-1)}{(l+1)(l+2)}}\sum_{m=-l}^{l} U^{lm}\mathcal{Y}_L^{lm} \, ,\\
    V_L&=-\frac{(l+1)!}{8l}\sqrt{\frac{2l(l-1)}{(l+1)(l+2)}}\sum_{m=-l}^{l} V^{lm}\mathcal{Y}_L^{lm} \, ;
\end{align}
\end{subequations}
see, e.g., Eq.~(2.10) of Ref.~\cite{Favata:2008yd}.
It is useful to make the definitions
\begin{subequations}
\begin{align}
    s_l \equiv {} & \frac{l!}{4}\sqrt{\frac{2l(l-1)}{(l+1)(l+2)}} \, ,\\
    g_l \equiv {} & -\frac{(l+1)!}{8l}\sqrt{\frac{2l(l-1)}{(l+1)(l+2)}} \, ,
\end{align}
\end{subequations}
though note that $s_l$ and $g_l$ should not be confused with $s^{l,(\pm)}_{l';l''}$ or $g^{l}_{l',m';l'',m''}$ defined in the main text.
By substituting the STF moments into Eq.~\eqref{eq:ang_mom_STF}, we can write $\delta J_i^{(\alpha=\beta=1)}$ as
\begin{align}
    \delta J_i^{(\alpha=\beta=1)} = {} & \sum_{l\geq 2} 
    (l+1)^2 \mu_{l+1} \sum_{m,m'}\left(b_l s_{l+1} g_l U_{l+1,m'} \bar{V}_{lm} \right.\nonumber\\
    &\left.-b_{l+1} s_l g_{l+1} \bar{U}_{lm} V_{l+1,m'}\right) \bar{\mathcal{Y}}^{lm}_L \mathcal{Y}^{l+1,m'}_{iL} \, .
\end{align}
We used the properties in Eqs.~\eqref{eq:conj_multipole_moments} and~\eqref{eq:conj_STF} to simplify the result. 
The quantity $\bar{\mathcal{Y}}^{lm}_L \mathcal{Y}^{l+1,m'}_{iL}$ can be written in terms of Clebsch-Gordan coefficients using Eq.~(2.26b) of~\cite{RevModPhys.52.299}, and it is only non-zero only when $m'$ satisfies $m'=m$ or $m'=m \pm 1$ (though note that we need to multiply the result in~\cite{RevModPhys.52.299} by a factor of $4\pi$ to account for the different normalization of the spherical harmonics used in~\cite{Compere:2019gft}). 
Evaluating the relevant Clebsch-Gordon coefficients gives 
\begin{align}
\label{eq:intrinsic_ang_mom_STF_interm}
    &\delta  J_i^{(\alpha=\beta=1)} = \sum_{l\geq 2,m} 
    \mu_{l+1} \frac{(l+1)(2l-1)!!}{l!}\sqrt{(2l+3)(2l+1)} \nonumber\\
    &\times \Big[\left(b_l s_{l+1} g_l U_{l+1,m} \bar{V}_{lm} -b_{l+1} s_l g_{l+1} \bar{U}_{lm} V_{l+1,m}\right)c_{lm}\xi^0_i \nonumber\\
    & + \left(b_l s_{l+1} g_l U_{l+1,m+1} \bar{V}_{lm} -b_{l+1} s_l g_{l+1} \bar{U}_{lm} V_{l+1,m+1}\right)b_{lm}^{(+)}\xi^1_i\nonumber\\
    & + \left(b_l s_{l+1} g_l U_{l+1,m-1} \bar{V}_{lm} -b_{l+1} s_l g_{l+1} \bar{U}_{lm} V_{l+1,m-1}\right)b_{lm}^{(-)}\xi^{-1}_i\Big] \, ,
\end{align}
where the basis vectors $\xi_i^0$ and $\xi_i^{\pm1}$ are defined in Eq.~(2.15) of~\cite{RevModPhys.52.299}:
\begin{align}
    \xi_i^0 = \delta_i^z \, , \qquad \xi_i^{\pm 1} = \frac 1{\sqrt 2} (\mp \delta_i^x - i \delta_i^y) \, .
\end{align}
To relate the multipole moments of the angular momentum to the components of the angular momentum in inertial Minkowski coordinates, we follow a procedure similar to that described in~\cite{Flanagan:2015pxa,Nichols:2017rqr}.
First we note that one can write the magnetic-parity vector harmonics as
\begin{equation}
    \bar{T}^{A}_{(b),1m} = \omega^i_{1m} \epsilon^{AB} D_B n_i \, ,
\end{equation}
where the $\omega^i_{1m}$ are then given by
\begin{subequations}
\label{eq:CartesianTo1m}
\begin{align}
    & \omega_{10}^{x}=0 \, , \qquad \omega_{10}^{y}=0 \, , \qquad \omega_{10}^{z}=\frac 12 \sqrt{\frac{3}{2\pi}} \, , \\
    & \omega_{1\pm1}^{x}=\mp \frac{1}{4}\sqrt{\frac{3}{\pi}} \, , \qquad \omega_{1\pm1}^{0y}=\frac{i}{4} \sqrt{\frac{3}{\pi}} \, , \qquad \omega_{1\pm1}^{0z}=0 \, .
\end{align}
\end{subequations}
Because the angular momentum is a linear functional of the vector field $Y^A$, then the relationship between $\delta J^{(\alpha=\beta=1)}_{1m}$ and $\delta J^{(\alpha=\beta=1)}_{i}$ is given by
\begin{equation} \label{eq:J1mFromJi}
    \delta J^{(\alpha=\beta=1)}_{1m} = \omega^i_{1m} \delta J^{(\alpha=\beta=1)}_{i} \, .
\end{equation}

After substituting Eq.~\eqref{eq:intrinsic_ang_mom_STF_interm} into Eq.~\eqref{eq:J1mFromJi}, we find that 
\begin{subequations}
\begin{align}
    \delta J_{10}^{(\alpha=\beta=1)} = {} & \frac{1}{16}\sqrt{\frac{3}{2\pi}}\sum\limits_{l\geq 2,m}
    a_l c_{lm}(\bar{U}_{lm}V_{l+1,m}-\bar{V}_{lm}U_{l+1,m}) 
    , ,\\
    \delta J_{1\pm1}^{(\alpha=\beta=1)} = {} & \frac{1}{32}\sqrt{\frac{3}{\pi}}\sum\limits_{l\geq 2,m}
    a_l b_{lm}^{(\pm)} (\bar{U}_{lm}V_{l+1,m\pm1}\nonumber\\
    &-\bar{V}_{lm}U_{l+1,m\pm1}) \, ,
\end{align}
\end{subequations}
where each term in the sum is a factor of $l+1$ larger than in Eq.~\eqref{eq:intrinsic_angular_momentum} as noted in the text after that equation.

We next perform a similar check for the center-of-mass angular momentum. 
Since only the $\beta$-dependent term was computed in~\cite{Compere:2019gft}, we convert their expression in terms of STF tensors and compare it to the $\beta$-dependent term in Eq.~\eqref{eq:CM_angular_momentum_w/out_parameters_condition}. 
We start from Eq.~(4.17) of~\cite{Compere:2019gft}, and we denote the second term by $\delta k_i^{(\beta=1)}$, which is given by
\begin{align}
    \delta k_i^{(\beta=1)} = {} & \sum_{l\geq 2} \Big[(l+1) \mu_{l+1}\left( U_{iL} U_L + b_l b_{l+1} V_{iL} V_L \right) \nonumber\\
    &+ \frac 12 \sigma_l \epsilon_{ijk} U_{jL-1}V_{kL-1}\Big] \, .
\end{align}
The coefficient $\sigma_l$ is defined in~\cite{Compere:2019gft} by
\begin{equation}
    \sigma_l = \frac{8(l+2)}{(l-1)(l+1)!(2l+1)!!} \, .
\end{equation}
We perform the same procedure of converting the $l$-index STF mass and current moments into the $U_{lm}$ and $V_{lm}$. 
The $\beta$-dependent difference term in the CM can then be written as follows:
\begin{align}
\label{eq:CM_ang_mom_STF_interm}
    \delta k_i^{(\beta=1)} = {} & \sum_{l\geq 2,m}\frac{(2l+1)!!}{l!}\Big\{\mu_{l+1} s_l s_{l+1} \sqrt{\frac{(2l+3)}{(2l+1)}}\nonumber\\
    &\times \Big[ (\bar{U}_{lm} U_{lm} + \bar{V}_{lm} V_{lm}) c_{lm} \xi^0_i \nonumber\\
    &+  (\bar{U}_{lm} U_{l,m+1} + \bar{V}_{lm} V_{l,m+1}) \frac{b^{(+)}_{lm}}{\sqrt{2}} \xi^1_i\nonumber\\
    &+(\bar{U}_{lm} U_{l,m-1} + \bar{V}_{lm} V_{l,m-1}) \frac{b^{(-)}_{lm}}{\sqrt{2}} \xi^{-1}_i\Big]\nonumber\\
    &+\frac{im}{2l}\sigma_l s_l g_l \bar{U}_{lm} V_{lm} \xi_i^0-\frac{d_{lm}^{(+)}}{\sqrt{2}} \bar{U}_{lm} V_{l,m+1} \xi_i^1\nonumber\\
    &+ \frac{d_{lm}^{(-)}}{\sqrt{2}} \bar{U}_{lm} V_{l,m-1} \xi_i^{-1}\Big\}
\end{align}
To relate the multipole moments of the CM angular momentum to its components in inertial Minkowski coordinates, we follow the same procedure as with the intrinsic angular momentum.
We first write the electric-type vector harmonics as
\begin{align}
    \bar{T}^{A}_{(e),1m} = \omega^{i}_{1m} D^A n_{i},
\end{align}
where the coefficients $\omega^{i}_{1m}$ are given in Eq.~\eqref{eq:CartesianTo1m}.
We can then solve for the multipole moments of the CM angular momentum given the relation
\begin{align}
\label{eq:CM_ang_mom_multipole_minkowski}
    \delta k_{1m}^{(\beta=1)} = \omega^{i}_{1m} \delta k_i^{(\beta=1)} \, .
\end{align}
Using Eqs.~\eqref{eq:CartesianTo1m} and~\eqref{eq:CM_ang_mom_multipole_minkowski} with Eq.~\eqref{eq:CM_ang_mom_STF_interm}, we find that the multipole moments of the CM angular momentum are
\begin{subequations}
\begin{align}
\delta k_{1,0}^{(\beta=1)}={}&-\frac{1}{16}\sqrt{\frac{3}{2\pi}}\sum\limits_{l\geq 2,m}\frac{1}{l+1}\Big[a_l c_{lm}\left(\bar{U}_{lm}U_{l+1,m}\right.\nonumber\\
&\left.+\bar{V}_{lm}V_{l+1,m}\right) -\frac{2im}{l}\bar{U}_{lm} V_{lm}\Big],\\
\delta k_{1,\pm1}^{(\beta=1)}={}&-\frac{1}{32}\sqrt{\frac{3}{\pi}}\sum\limits_{l\geq 2,m}\frac{1}{l+1}\Big[a_l b_{lm}^{(\pm)}\left(\bar{U}_{lm}U_{l+1,m\pm1} \right.\nonumber\\ 
&\left.+\bar{V}_{lm}V_{l+1,m\pm1}\right) \pm \frac{2i}{l}d_{lm}^{(\pm)}\bar{U}_{lm} V_{l,m\pm1}\Big].
\end{align}
\end{subequations}
This is identical to the result in Eq.~\eqref{eq:CM_angular_momentum_w/out_parameters_condition}.

\bibliography{refs}

\end{document}